\documentclass[twocolumn]{aa}  
\usepackage{graphicx}
\usepackage{color}
\usepackage{epsfig}
\usepackage{supertabular}
\usepackage{aalongtable}
\usepackage{rotating}
\usepackage{txfonts}
\begin{document}
\title{The Nature of UCDs: Internal Dynamics from an Expanded Sample and Homogeneous Database\thanks{Based on observations obtained in service
    mode at the ESO Paranal Observatory with the VLT (programme 078.B-0496)}}

\titlerunning{The Nature of UCDs}

    \author{Steffen Mieske
          \inst{1}
          \and
          Michael Hilker\inst{1}\and Andr\'{e}s Jord\'{a}n\inst{2,3} \and Leopoldo Infante\inst{3} \and Markus Kissler-Patig\inst{1} \and Marina Rejkuba\inst{1} \and Tom Richtler\inst{4} \and Patrick C\^{o}t\'{e}\inst{5} \and Holger Baumgardt\inst{6} \and Michael J. West\inst{7} \and Laura Ferrarese \inst{5} \and Eric W. Peng\inst{5}
          }

   \offprints{S. Mieske}

   \institute{European Southern Observatory, Karl-Schwarzschild-Strasse 2, 85748 Garching bei M\"unchen, Germany\\
              \email{smieske@eso.org}
         \and
     Clay Fellow, Harvard-Smithsonian Center for Astrophysics,
60 Garden St.,
Cambridge, MA 02138, USA
\and
     Departamento de Astronom\'{\i}a y Astrof\'{\i}sica, Pontificia
Universidad Cat\'olica de Chile, Casilla 306, Santiago 22, Chile
\and
     Universidad de Concepci\'{o}n, Departamento de Fisica, Astronomy Group, Casilla 160-C, Concepci\'{o}n, Chile
\and
      National Research Council of Canada,
Herzberg Institute of Astrophysics,
5071 West Saanich Road,
Victoria, BC V9E 2E7, Canada
\and
      Argelander-Institut f\"ur Astronomie, Auf dem H\"ugel 71, 53121 Bonn, Germany\and
      European Southern Observatory, Alonso de C\'{o}rdova 3107,
Vitacura, Casilla 19001, Santiago, Chile}

   \date{}

 
  \abstract
   {The internal dynamics of ultra-compact dwarf galaxies (UCDs) has attracted increasing attention, with most of the UCDs studied to date located in the Virgo cluster.} 
{Our aim is to perform a comprehensive census of the internal dynamics
  of UCDs in the Fornax cluster, and to shed light on the nature of the interface between star clusters and galaxies. }
{We obtained high-resolution spectra of 23 Fornax UCDs
  with $-10.4>M_V>-13.5$ mag ($10^6<{\rm M/M_{\sun}<10^8}$), using
  FLAMES/Giraffe at the VLT. This is the largest homogeneous data set of UCD internal dynamics assembled to date. We derive dynamical M/L ratios for 
  15 UCDs covered by HST imaging.}
{In the M$_V$-$\sigma$ plane, UCDs with $M_V<-12$ mag are consistent
  with the extrapolated Faber-Jackson relation for luminous elliptical
  galaxies, while most of the fainter UCDs are closer to the
  extrapolated globular cluster (GC) relation. At a given metallicity,
  Fornax UCDs have, on average, M/L ratios lower by 30-40\% than Virgo
  UCDs, suggesting possible differences in age or dark matter content
  between Fornax and Virgo UCDs.  For our sample of Fornax UCDs we
  find no significant correlation between M/L ratio and mass.  We
  combine our data with available M/L ratio measurements of compact
  stellar systems with $10^4<{\rm M/M_{\sun}}<10^8$M, and normalise
  all M/L estimates to solar metallicity.  We find that UCDs
  (M$\gtrsim$2$\times 10^6$M$_{\sun}$) have M/L ratios twice as large
  as GCs (M$\lesssim$2$\times 10^6$M$_{\sun}$). We argue that
  dynamical evolution has probably had only a small effect on the
  current M/L ratios of objects in the combined sample, implying that
  stellar population models tend to under-predict dynamical M/L ratios
  of UCDs and over-predict those of GCs. Considering the scaling relations
  of stellar spheroids, we find that UCDs align well along the 'Fundamental
  Manifold'.  UCDs can be considered the small-scale end of the galaxy
  sequence in this context. The alignment for UCDs is especially clear
  for $r_e\gtrsim 7$pc, which corresponds to dynamical relaxation
  times that exceed a Hubble time. In contrast, globular clusters exhibit a
  broader scatter and do not appear to align along the manifold.}
{We argue that UCDs are the smallest dynamically un-relaxed
  stellar systems, with M$\gtrsim$2$\times 10^6$M$_{\sun}$ and
  7$\lesssim$${\rm r_e/pc}$$\lesssim$100. Future studies should aim at
  explaining the elevated M/L ratios of UCDs and the environmental
  dependence of their properties.}

   \keywords{galaxies: clusters: individual: Fornax -- galaxies:
dwarf -- galaxies: fundamental parameters -- galaxies: nuclei --
galaxies: star clusters}

   \maketitle 
%

\section{Introduction}
\label{intro}
In recent years, significant effort has been devoted to studying the
internal dynamics of extragalactic compact stellar systems in the mass
regime of massive globular clusters and ultra-compact
dwarf galaxies ($10^6<M/M_{\sun}<10^8$) (Drinkwater et al. 2003, Martini \& Ho 2004, Ha\c{s}egan et al. 2005, Maraston
et al. 2004, Rejkuba et al. 2007, Evstigneeva et
al. 2007, Hilker et al. 2007). A compilation of the available data is
presented in Mieske \& Kroupa (2008) and Dabringhausen, Hilker \&
Kroupa (2008).

A striking outcome of these studies is the finding that the dynamical M/L ratios
of massive compact stellar systems are, on average, about two times
larger than those of normal globular clusters of comparable metallicity.
Several objects have M/L ratios at the limit of, or even beyond, the range
predicted by stellar population models assuming canonical IMFs
(Ha\c{s}egan et al. 2005).  Possible explanations for these high M/L ratios
include extreme stellar mass functions (Mieske \& Kroupa 2008;
Dabringhausen, Hilker \& Kroupa 2008) or densely packed dark matter
(Goerdt et al. 2008).  The occurence of objects with high M/L ratios
is observed to start at $\sim 2\times$10$^6$M$_{\sun}$ (Ha\c{s}egan et
al.  2005, Rejkuba et al.  2007, Mieske \& Kroupa 2008), coinciding
with ``breaks" in physical size (Ha\c{s}egan et al. 2005, Mieske
et al.  2006, Kissler-Patig et al.  2006) and stellar
content (Mieske et al. 2006).  These breaks are consistent with the
hypothesis that at $\sim 2\times$10$^6$M$_{\sun}$ ($M_V \simeq -11$
mag) we observe the transition between simple globular clusters and
more complex systems, the UCDs.

Three main formation scenarios have been suggested for UCDs: 1. UCDs
are stellar super clusters formed in the tidal arms of violent
gas-rich galaxy mergers (Fellhauer \& Kroupa 2002, 2005). 2. UCDs are
tidally stripped compact remnants of nucleated dwarf galaxies (Bassino
et al. 1994, Hilker et al. 1999, Bekki et al.  2003, Goerdt et al.
2008). 3. UCDs are genuine compact dwarf galaxies formed from
small-scale peaks in the primordial dark matter power spectrum
(Drinkwater et al. 2004). In the first case, UCDs are not expected to
contain any dark matter. In the second and third case, UCDs would be
related to cosmological low-mass dark matter halos and may contain
dark matter (Goerdt et al.  2008). The comparably large M/L ratios of
5-10 found for some Virgo UCDs may point towards a cosmological
origin. These M/L ratios are similar to the values found for some of
the more luminous Local Group dSphs like Sculptor and LeoI (Gilmore et
al. 2007), although note that they are still 1-2 orders of magnitude
below the M/L values found for the ultra-faint dSph candidates (Gilmore et al.
2007, Simon \& Geha 2007).

In this paper, we aim to study whether the high M/L ratios are a
fundamental trend equally common to all UCDs, or whether environmental
variations of the trend exist.  The latter may be expected if competing
formation channels dominate in different environments (e.g. Mieske et
al. 2006). For both the Virgo cluster (Ha\c{s}egan et al. 2005,
Evstigneeva et al.  2007) and the Centaurus A group (Rejkuba et al.
2007), more than 10 objects in the UCD mass range have measured M/L
ratios. For the Fornax cluster, only five sources have high resolution
spectroscopy available (Hilker et al.  2007), making it difficult
to judge whether differences exist between Fornax and
Virgo UCDs. With a comprehensive sample of measurements for Fornax, it
will be possible to analyse whether Fornax UCDs extend to such high
M/L ratios as Virgo UCDs.

In what follows, we present new measurements of the internal kinematics of 23
compact objects in Fornax (obtained with FLAMES/GIRAFFE at the VLT). We
analyse how their M/L ratios relate to predictions from stellar
population models, and investigate how they fit into the trend of
increasing M/L with mass among compact stellar systems. We also
examine how UCDs and globular
clusters fit into the broader context of larger and more luminous
stellar systems, focusing on the fundamental manifold of stellar spheroids
(Zaritsky et al. 2006a, 2006b, 2008) that is an extension of the
fundamental plane concept.  

The paper is structured as follows: in Sect.~\ref{data} we present the
new spectroscopic data of Fornax UCDs used for the present study.
Section~\ref{reduction} describes the data reduction, including the
modelling of the mass distribution. In Sect.~\ref{results}, the
results for the Fornax UCDs are presented and discussed. In
Sect.~\ref{discussion} we combine the Fornax data with other
literature results on dynamical M/L ratios of compact stellar systems,
and investigate UCDs and GCs in the context of the fundamental
manifold. The paper finishes with Summary and Conclusions in
Sect.~\ref{conclusions}.  Throughout this paper we assume a distance
modulus to Fornax of (m-M)=31.4 mag (Freedman et al.  2001).

\section{Data}
\label{data}
The data for this study were obtained in service mode with the
Fibre Large Array Multi Element Spectrograph (FLAMES; Pasquini et
al.~2002) mounted on UT2 at the VLT (programme 078.B-0496).  We used
the spectrograph GIRAFFE in MEDUSA mode, which allows the observation
of up to 130 targets at the same time over a 25 arcmin diameter field of view,
using fibres of 1.2$''$ aperture. We observed a total of 15 hours
on-source, subdivided in 15 individual integrations of 1 hour duration.

\begin{figure*}
\begin{center}
  \epsfig{figure=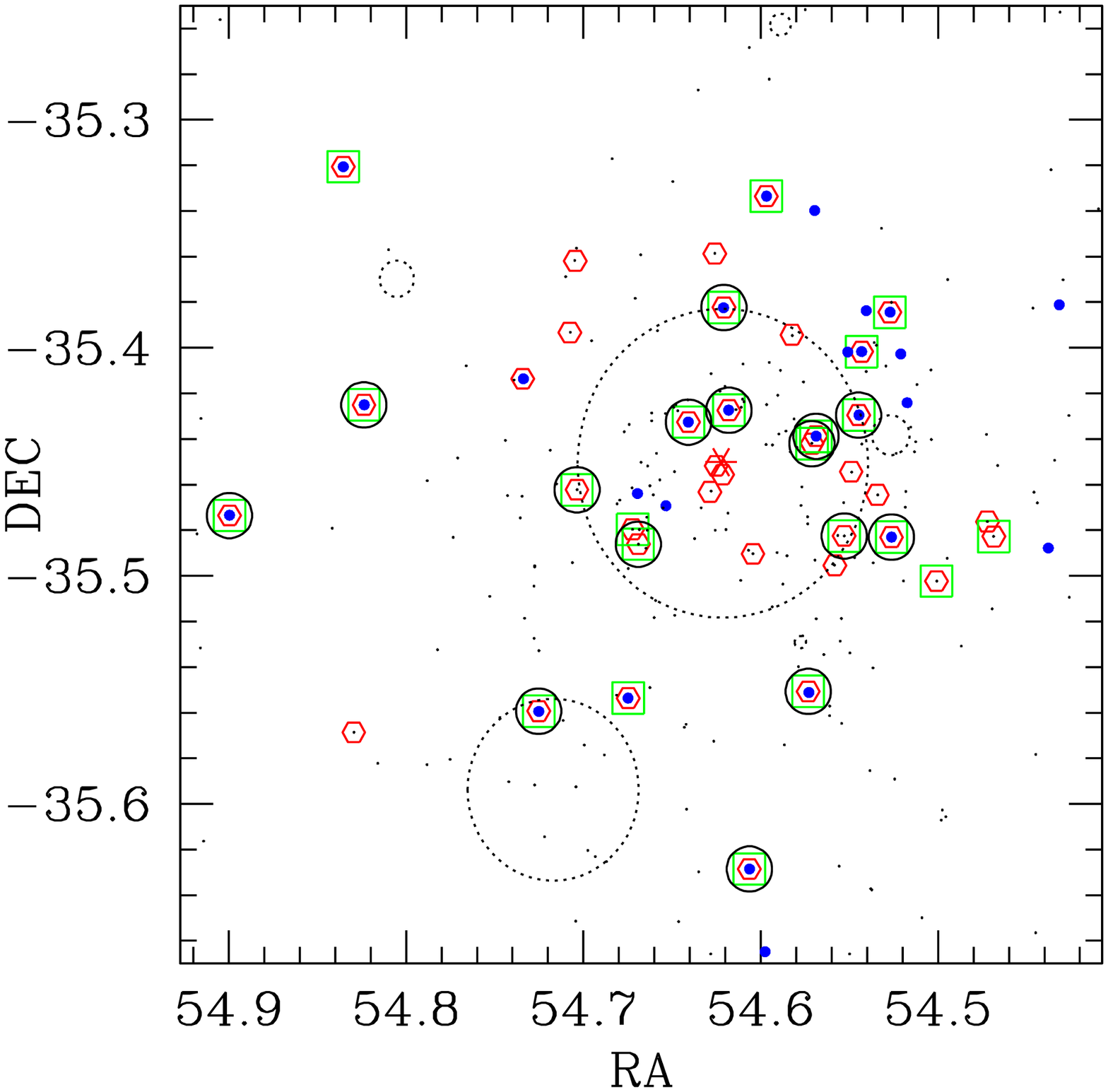,width=8.6cm}
  \epsfig{figure=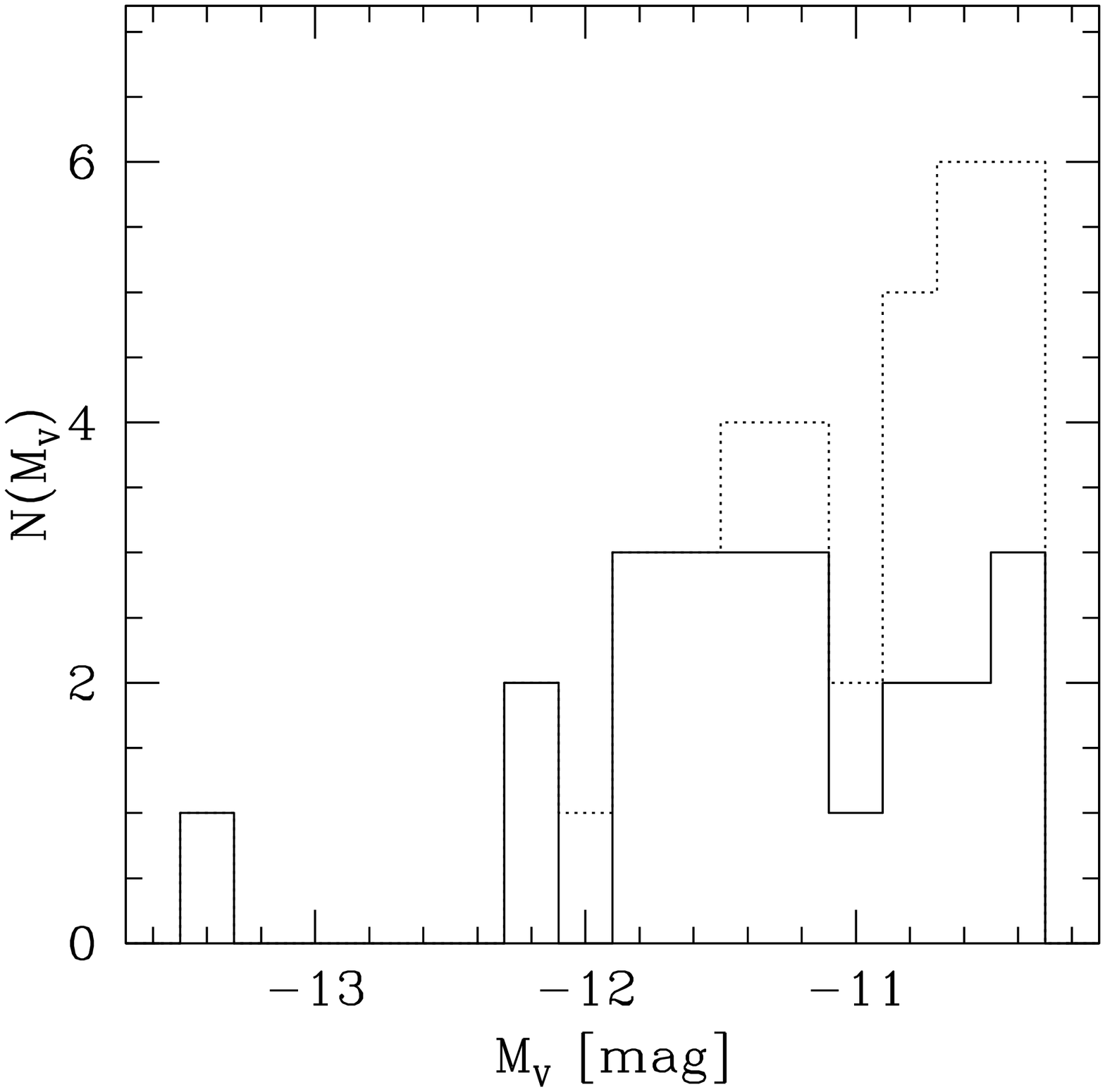,width=8.6cm}
  \caption{{\bf Left:} Map of the observed region in the Fornax cluster.  Hexagons indicate
    compact objects observed with FLAMES. Green squares
    indicate objects for which reliable velocity dispersions were measured.
    Large circles show targets with HST
    imaging that were successfully observed with FLAMES.  Small dots are all known compact cluster members with
    $V<22$ mag ($M_V<-9.4$ mag), large dots are those with $M_V<-11$
    mag, the approximate magnitude division between UCDs and GCs.  The dotted
    circles indicate Fornax cluster member galaxies from the Fornax
    Cluster Catalog (FCC; Ferguson 1989), for which the circle size
    gives the radius at which $\mu_V=25$ mag / arcsec$^2$. The
    asterisk marks the location of NGC 1399. {\bf Right:}
    Magnitude distribution of observed sources. The solid histogram refers to those
    objects for which reliable velocity dispersions were measured.
    The dotted histogram refers to all sources included in the fibre
    configuration (see the left panel).}

\label{map}
\end{center}
\end{figure*}

Fig.~\ref{map} shows a map of the observed region. We observed 37
compact objects within 12$'$ of NGC 1399 and with $18<V<21$ mag
($-13.4<M_V<-10.4$ mag). This magnitude range covers the UCDs and
overlaps the bright end of the globular cluster luminosity function
(Mieske et al. 2004). All targets have confirmed cluster membership
from spectroscopic surveys (Drinkwater et al. 2000, Mieske et al. 2002
\& 2004, Richtler et al.  2004 \& 2008), except for the two objects
closest to NGC 1399, which were selected on the basis of their morphology
from imaging from the ACS Fornax cluster
survey (Jord\'an et al. 2007).

For the observations we used the HR09A grism, which provides an
instrumental resolution of 8 km/s in terms of Gaussian $\sigma$ (or 19
km/s in terms of FWHM) over a wavelength range $5100 < \lambda < 5400$
\AA.  This resolution allows us to reliably measure velocity dispersions
$\ge$ 10 km/s.

\section{Data reduction}
\label{reduction}

\subsection{Basic reduction}
In order to remove the instrumental signatures from the data, we used
the publicly available GIRAFFE data reduction pipeline from
ESO\footnote{http://www.eso.org/projects/dfs/dfs-shared/web/vlt/vlt-instrument-pipelines.html}.
This pipeline performs bias subtraction, flat-field division,
wavelength calibration, and spectrum extraction. As such, it creates a
wavelength calibrated 1D spectrum from a raw 2D spectrum. The pixel
scale in the wavelength calibrated 1D spectrum is 0.05 {\AA} per
pixel, slightly over-sampling the instrumental scale of 0.08 {\AA} per
pixel. The RMS of the wavelength solution was of the order 0.1 {\AA}.
The instrumental resolution --- resulting from the four pixel FWHM of
the fiber's spatial profile --- in terms of Gaussian $\sigma$ is
$\sim$0.14 {\AA}. This instrumental resolution corresponds to a velocity
dispersion of $\sigma$$\sim$8 km/s in the 5100--5400 \AA\ wavelength
regime.

Given the multiplexing capability of FLAMES/GIRAFFE, we also obtained
23 sky spectra in each exposure by assigning unoccupied fibres to
empty sky positions. These spectra were reduced identically to the
science spectra. We combined the 23 sky spectra in each exposure to 1
master sky spectrum, using the IRAF task \texttt{scombine} in the ONEDSPEC
package. This combined sky spectrum was subtracted from each single,
calibrated 1D object spectrum.

From this we obtained 15 sky-subtracted, calibrated 1D spectra for
each of the 37 compact objects observed. Those single spectra were
corrected to heliocentric velocity using the IRAF tasks \texttt{rvcorrect} in
the RV package and \texttt{dopcor} in the ONEDSPEC package. The velocity shift
between the 15 individual spectra due to shifts in wavelength calibration
was very small ($<2$ km/s). We combined the 15 registered single
spectra using the IRAF task \texttt{scombine}. For this we normalised the
intensity of the spectra to their mode and applied a 3.5$\sigma$
average sigma clipping algorithm. The resulting S/N per pixel in the combined
object spectra ranged between 5 and 35.

In order to have template spectra for measuring the internal velocity
dispersion, we also observed  several dozen red giant
stars in the Milky Way globular cluster $\omega$ Centauri in a single FLAMES/GIRAFFE
pointing. These stars are of late spectral type (typical temperature 5000
K), cover a metallicity range $-2.3<{\rm [Fe/H]}<-0.5$ dex and have
magnitudes around $V=12$ mag (van Loon et al. 2007). We used the same
instrument setting and reduction procedures as for the science
targets. The internal line width of the giant stars is negligible
compared to the instrumental resolution. With 5 minutes on-source integration, we
reached S/N ratios between 50 and 100. We used the 14 highest quality spectra 
as templates for the dispersion measurements.

\begin{figure}
\begin{center}
  \epsfig{figure=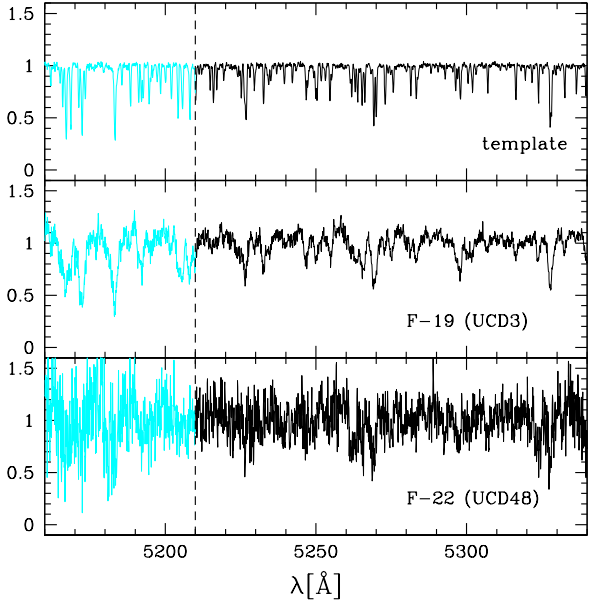,width=8.6cm}
  \caption{Continuum normalised spectra of one template and two science objects, shifted to the restframe. The wavelength region $\lambda_{\rm
      restframe}<5210 \AA$ containing the Mg features was excluded
    from the cross-correlation fit (see Fig.~\ref{compportion} and
    text). }
\label{spectra}
\end{center}
\end{figure}

In Fig.~\ref{spectra} we show examples of two object spectra
and one template spectrum.

\subsection{Dispersion measurements}
\begin{figure}
\begin{center}
  \epsfig{figure=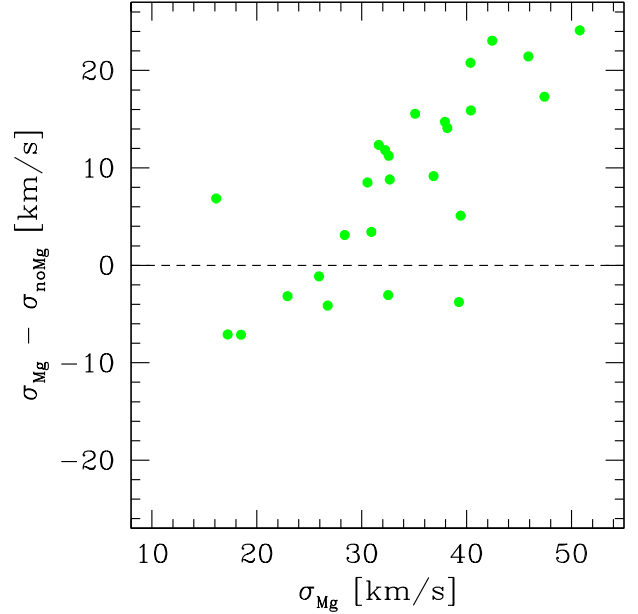,width=8.6cm}
  \caption{The x-axis shows the Gaussian width $\sigma_{\rm Mg}$ in km/s of the cross-correlation peak derived in the Mg region ($5170<\lambda_{\rm restframe}<5210$ $\AA$). The y-axis shows the difference between this width and $\sigma$ derived in the rest of the spectrum ($5210<\lambda_{\rm
      restframe}<5390$ $\AA$). For $\sigma_{\rm Mg} \gtrsim 30$ km/s, the cross
    correlation width in the Mg region is significantly broader than
    in the rest of the spectrum.}
\label{compportion}
\end{center}
\end{figure}

The internal velocity dispersion, $\sigma$, of each compact object was
measured by cross-correlating its spectrum with various template spectra
(IRAF task \texttt{fxcor}; Tonry \& Davis~1979). For this measurement
we excluded the wavelength region around the very strong Mgb lines
($\lambda_{\rm restframe}<5210$ $\AA$), since the measured width
in this region proved to be systematically larger than in the rest of
the spectra (see Fig.~\ref{compportion}). Such an increased width in
the very deep $\alpha$ element absorption features is likely caused by
saturation effects, and has been found previously by other authors
(e.g., Rejkuba et al.~2007 and Hilker et al.~2007), who also excluded
this region from their measurements. We used the wavelength region
$5210<\lambda_{\rm restframe}<5390$ $\AA$, which includes the
many prominent Fe absorption features around 5325 $\AA$ (see
Fig.~\ref{spectra}).

\begin{figure*}
\begin{center}
  \epsfig{figure=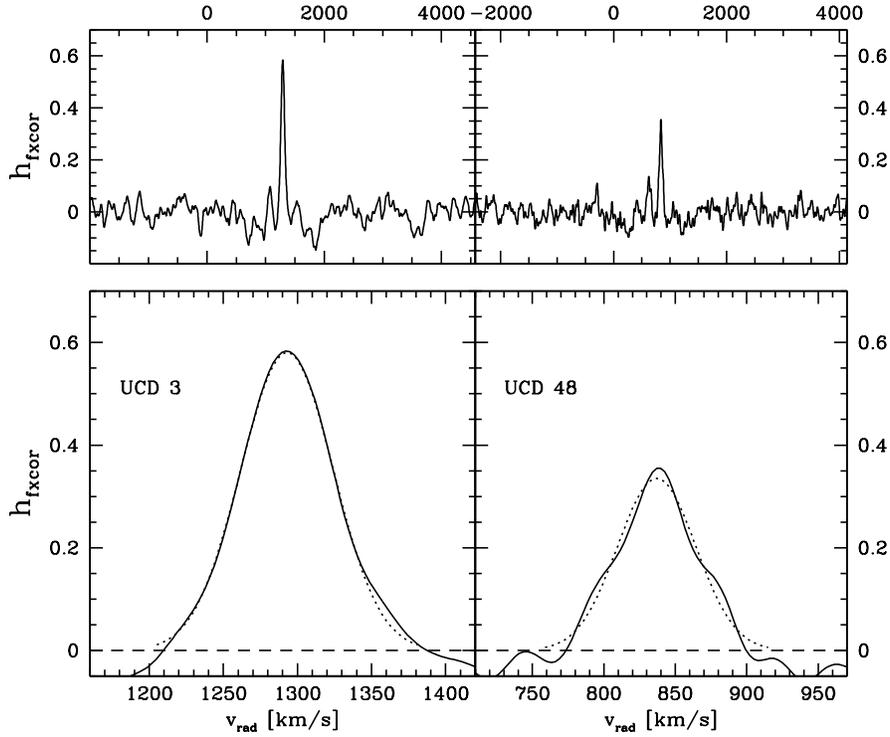,width=12cm}
  \caption{Plot showing the Fourier cross-correlation peaks for objects F-${19}$ (UCD3, left) and F-${22}$ (UCD48, right), whose spectra are shown in Fig.~\ref{spectra}. The lower panels  are magnified views of the overall cross-correlation results from the upper panels. In the lower panels, the Gaussian fit to the cross-correlation peak is indicated by a dotted line. See text for more details. }
\label{fxcor}
\end{center}
\end{figure*}

Prior to cross-correlation, we continuum subtracted the spectra. For
this, we adjusted the continuum fitting order individually for each
source such as to yield the lowest order that gives satisfactory
results.  The peak position of the cross-correlation gives the
relative radial velocity between object and template. The width,
$\sigma_{\rm peak}$, of the cross-correlation peak (Fig.~\ref{fxcor})
is the quadratic sum of the intrinsic object line width caused by
random stellar motion plus twice the instrumental line width (equal to
the template line width): i.e., $\sigma_{\rm peak}^2=\sigma_{\rm
  obj}^2+2\times \sigma_{\rm ins}^2$.  By cross-correlating the
un-broadened and continuum subtracted templates against each other, we
measured the template's intrinsic line width $\sigma_{\rm ins}$ to be
9.7 km/s with a very small scatter of order 0.4 km/s.  The intrinsic
line width $\sigma_{\rm obj}$ of the object spectrum is then
calculated as: $\sigma_{\rm obj}=\sqrt{\sigma_{\rm peak}^2-2 \times
  \sigma_{\rm ins}^2}$. Note that the factor 2 in front of $\sigma_{\rm ins}^2$
is necessary because both the object and template spectrum are
broadened by the instrumental resolution (Dubath et al. 1992).  In
Fig.~\ref{fxcor}, we show the cross-correlation results from the two
objects whose spectra are shown in Fig.~\ref{spectra}.

\begin{figure}[]
\begin{center}
  \epsfig{figure=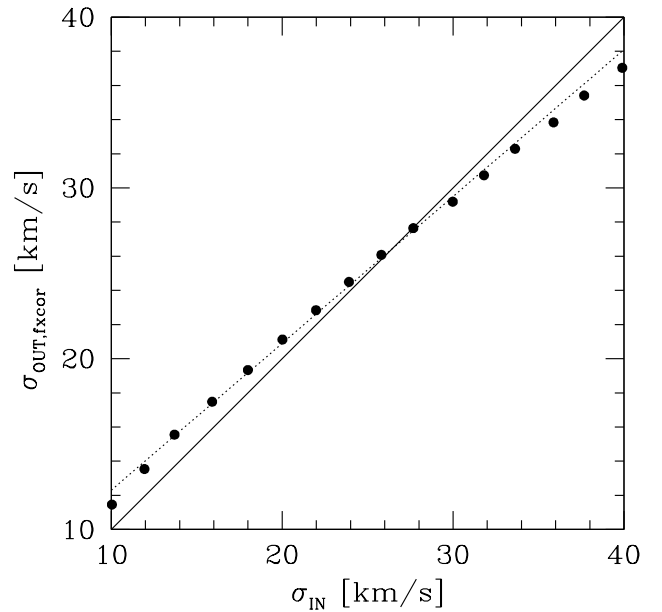,width=8.6cm}
  \caption{Plot illustrating the tests performed with cross-correlating
    artificially broadened spectra. The input width, $\sigma_{\rm IN}$, is
    plotted against the width, $\sigma_{\rm OUT}$, measured using the \texttt{fxcor} task.
    The solid line shows the identity relation. The dotted line is a linear fit
    to $\sigma_{\rm OUT}$ as a function of $\sigma_{\rm IN}$. The measured values of $\sigma$ are corrected for according to this relation.
    The values shown here are for the case of k=3 as low frequency
    cut-off. For $k>3$, the deviations from a linear relation were
    larger.}
\label{dispcor}
\end{center}
\end{figure}

We performed tests with artificially broadened template spectra and
different low frequency Fourier filter cutoffs to assess the accuracy
of the \texttt{fxcor} task in measuring $\sigma_{\rm peak}$. We found a
slightly non-linear relation between input spectral width and width
measured by \texttt{fxcor}. The best agreement was found for a low-frequency
cutoff of k=3 (see Fig.~\ref{dispcor}).  We adopted this cut-off for the
Fourier filtering, and applied a residual correction as a linear
function of $\sigma_{\rm obj}$ --- as indicated in Fig.~\ref{dispcor}
--- to the measured width of the science spectra.  The residual
correction is independent of the S/N in the object spectra, which we
tested by artificially degrading the broadened template spectra to a
range of S/N values between 5 and 35 per pixel, representative for our
compact object sample. From the tests with the template spectra we
also found that the background value in the cross-correlation peak fit
needs to be kept fixed at 0 (see also Fig.~\ref{fxcor}).  Allowing the
program to fit the background value led to consistently over-estimated
widths.

We accepted a reliable measurement of $\sigma_{\rm obj}$ for a given
object if two conditions were met: (1) the average confidence level of the
cross-correlation peak was $R>4$; and (2) none of the template
cross-correlations yielded an outlier in the template-object relative
velocity. The first condition removed 9 sources from the main sample
of 37 objects while the second condition removed five more sources.
Fig.~\ref{map} shows that the rejected sources are mostly close to the
faint magnitude limit of our survey. We note that the two brightest
sources with unreliable measurements (see also Fig.~\ref{map}) are
those that had been selected as UCD candidates based only on
morphology from ACS imaging (ACS Fornax cluster survey, see Jord\'an
et al. 2007). Both sources are located within 2$'$ to the center of
NGC 1399 and are the only objects in our target sample whose
coordinates could not be tied to the USNO B2.0 system. We
attribute their low flux level to an offset in relative
coordinates with respect to the rest of our sample.

A final sample of 23 reliable measurements is obtained, of which 15
have archival HST imaging available. {\it This is the largest
  homogeneous set of UCDs for which dynamical masses have been
  derived.}  The resulting range of intrinsic velocity dispersions is
$9<\sigma_{\rm obj}<36$ km/s, with a mean of 24 km/s. These values are
listed in Table~\ref{table1} for the 15 sources with HST imaging, and
in Table~\ref{table1a} for the 8 sources without HST imaging. A map
and the magnitude distribution of the investigated compact objects is
shown in Fig.~\ref{map}.

\begin{sidewaystable}
\caption{Table showing the measured velocity dispersions, $\sigma_{\rm obj}$, for the Fornax
  UCDs with available HST imaging, plus the true global and central
  velocity dispersions $\sigma_{\rm obj,cor}$ and $\sigma_{\rm 0}$,
  derived from modeling the light distribution of these sources within
  the FLAMES fibre aperture (see text and Hilker et al. 2007 for
  further details). We also indicate the derived total mass, the
  optical mass-to-light ratio M/L$_V$, and the effective radius
  r$_{\rm eff}$. Coordinates are taken from USNO. The first error of
  M/L$_V$ corresponds to the measurement uncertainty including that of
  the apparent magnitude. The second one is the systematic error
  arising from an assumed 0.15 mag uncertainty in the distance modulus
  of (m-M)=31.4 mag to Fornax. The heliocentric radial velocities have
  precisions of $\sim$3 km/s, and a global uncertainty of 1-2 km/s,
  tied to the heliocentric velocity of $\omega$ Centauri (233 km/s,
  Dinescu et al. 1999) . The fourth-to-last column shows the dynamical
  M/L derived by Hilker et al. (2007) for the three brightest sources.
  The three last columns indicate the relaxation time t$_{\rm relax}$,
  the acceleration parameter a, and the projected distance d to NGC
  1399 (1$'$=5.5 kpc at the Fornax cluster distance). $^*$Designation
  from Drinkwater et al. (2003).  $^{**}$Literature ID refers to
  Mieske et al. (2004) for FCOS, and Firth et al. (2007) for the
  UCDxx$_{\rm DW}$ sources.}
\label{table1}
\begin{center}
\tiny
\begin{tabular}{|ll|rrrrrrrrrrr|r|rrr|}
\hline
ID$_{\rm FLAMES}$ & ID$_{\rm literature}^{**}$ &RA [2000] & DEC [2000]& V$_0$       & M$_V$       & v$_{\rm rad,\sun}$       &$\sigma_{\rm obj}$        & $\sigma_{\rm obj,cor}$        & $\sigma_{\rm 0}$        & Mass                     & M/L                         & r$_{\rm eff}$  &   M/L$_{\rm H07}$ & t$_{\rm relax}$ & a & d\\
                  &                         &          &           &   [mag] &       [mag] &                    [km/s]&                   [km/s] &                        [km/s] &                  [km/s] &      [10$^6$ M$_{\sun}$] &     [M$_{\sun}$/L$_{\sun}$] &               [pc] & [M$_{\sun}$/L$_{\sun}$] & [t$_{\rm Hubble}$] & [m s$^{-2}$] & [']\\\hline

F-${19}$ (UCD3)$^*$ & FCOS 1-2053 &  3:38:54.1 & -35:33:33.6 &    18.0 & -13.4 & 1509 & 27.5 (1.7) &  22.8 &  29.5 &  93.6 (15.0) &  4.7 (0.7) (0.35) &  89.7 & 4.1 (1.0) & 263 & 1.6E-9 & 8.2\\ 
F-${24}$ (UCD4)$^*$ & FCOS 1-2083 &  3:39:35.9 & -35:28:24.6 &  19.1 & -12.3 & 1902 & 24.5 (1.7) &  21.4 &  30.3 &  24.5 (4.3) &  3.4 (1.1) (0.26) &  29.5 & 4.6 (1.1) & 27.3 & 3.9E-9  & 13.5\\ 
F-${1}$ (UCD2)$^*$ & FCOS 2-2111 &  3:38:06.3 & -35:28:58.8 &  19.2 & -12.2 & 1261 & 21.3 (1.7) &  18.7 &  27.1 &  16.2 (3.2) &  2.5 (0.6) (0.18) &  23.1 & 3.1 (0.5) & 15.8 & 4.2E-9  & 5.1\\ 
F-${5}$ & FCOS 2-2134 &  3:38:10.8 & -35:25:46.2 &  19.7 & -11.7 & 1672 & 35.6 (2.9) &  34.5 &  43.5 &  13.7 (2.7) &  3.2 (0.6) (0.24) &  5.0 &  & 1.48 & 7.6E-8  & 4.0\\ 
F-${12}$ & FCOS 0-2031 &  3:38:29.0 & -35:22:56.3 &  19.9 & -11.5 & 1661 & 23.9 (1.9) &  22.9 &    30.0 &  8.3 (1.6) &  2.4 (0.8) (0.18) &  10.3 &  & 3.51 & 1.1E-8  &  4.1\\ 
F-${11}$ & FCOS 0-2030 &  3:38:28.3 & -35:25:38.5 &  19.9 & -11.5 & 1706 & 27.1 (1.8) &  26.2 &  32.3 &  5.7 (1.0) &  1.6 (1.1) (0.12) &   3.6 &  & 0.62 & 6.2E-8 & 1.4\\ 
F-${9}$ & FCOS 1-2024 &  3:38:25.5 & -35:37:42.7 &  20.1 & -11.3 & 1752 & 30.9 (2.0) &  25.7 &  31.2 &  14.1 (2.4) &  4.7 (1.2) (0.35) &  9.1 &  & 3.64 & 2.4E-8  & 10.8\\ 
F-${17}$ & FCOS 1-2095 &  3:38:33.8 & -35:25:57.2 &  20.1 & -11.3 & 1390 & 27.7 (1.8) &  28.5 &  35.1 &  6.3 (1.0) &  2.2 (0.6) (0.17) &   3.3 &  & 0.56 & 8.0E-8  & 1.4\\ 
F-${7}$ & UCD33$_{\rm DW}$ &  3:38:17.6 & -35:33:02.4 &  20.3 & -11.1 & 1520 & 19.4 (1.8) &  20.1 &  24.1 &  10.5 (2.3) &  4.2 (0.6) (0.32) &  14.9 &  & 6.76 & 6.6E-9  & 6.5\\ 
F-${22}$ & UCD48$_{\rm DW}$&  3:39:17.7 & -35:25:30.1 &  20.3 & -11.1 & 1054 & 23.2 (1.7) &  22.8 &    32.0 &  5.3 (1.0) &  2.1 (0.6) (0.16) &    10.0 &  & 2.75 & 7.3E-9  & 9.9\\ 
F-${6}$ & FCOS 0-2024 &  3:38:16.5 & -35:26:19.6 &  20.3 & -11.1 &  857 & 25.6 (1.9) &  27.3 &  33.7 &  12.5 (2.3) &  5.3 (1.0) (0.40) &  7.3 &  & 2.49 & 3.3E-8  & 2.7\\ 
F-${34}$ & FCOS 0-2023 &  3:38:12.7 & -35:28:57.0 &  20.7 & -10.7 & 1639 & 26.1 (1.8) &  24.6 &  29.8 &  5.5 (1.0) &  3.2 (0.8) (0.24) &  4.0 &  & 0.69 &  4.9E-8 & 3.9\\ 
F-${51}$ & FCOS 0-2089 &  3:38:17.1 & -35:26:31.3 &  20.8 & -10.6 & 1257 & 19.5 (1.7) &  20.1 &  27.2 &  3.5 (0.7) &  2.4 (0.6) (0.18) &  4.2 &  & 0.62 & 2.8E-8 & 2.5\\ 
F-${53}$ & FCOS 1-2077 &  3:38:40.6 & -35:29:10.0 &  20.8 & -10.6 &  681 & 20.4 (1.8) &  19.6 &  24.8 &  3.9 (0.8) &  2.7 (0.7) (0.20) &  4.5 &  & 0.71 & 2.7E-8  & 3.1\\ 
F-${59}$ & FCOS 1-2089 &  3:38:48.9 & -35:27:44.1 &  20.9 & -10.5 & 1559 & 9.3 (1.8) &   9.8 &    12.0 &  1.3 (0.6) & 0.9 (0.4) (0.07) &   5.7 &  & 0.64 & 5.5E-9  & 4.0\\\hline

\end{tabular}
\normalsize
\end{center}
\end{sidewaystable}

\begin{table*}
\caption{Columns 2 to 6 of this table show the structural parameters of the King profile fits for the 15 UCDs from Table~\ref{table1} (see also Evstigneeva et al. 2008). The King profiles are parameterized as follows: $I(r) = I_0 \left[\frac{1}{(1+(r/r_c)^2)^{\frac{1}{\alpha}}} - \frac{1}{(1+(r_t/r_c)^2)^{\frac{1}{\alpha}}} \right]^{\alpha}$. For the brightest source (F-19), a composite King+Sersic profile was necessary to provide a satisfactory fit. The Sersic profile is parametrized as follows: $I(r) = I_{\rm eff} \,\, exp \left[-k \left(\left(\frac{r}{r_{\rm eff}}\right)^{\frac{1}{n}} - 1\right) \right]$. The parameters of the Sersic profile are indicated in columns 7 to 9.}
\begin{center}
\begin{tabular}{|l|rrrrr|rrr|}\hline
ID$_{\rm FLAMES}$ & $\mu_0$ [mag/arcsec$^2$]& r$_c$ [pc] & r$_t$ [pc]& c &  $\alpha$ & $\mu_{\rm eff}$ [mag/arcsec$^2$] & r$_{\rm eff}$ [pc] & n \\\hline
F-${19}$ &16.03 &4.90 &230.6 &1.67& 2& 21.34 &118.9&1\\
F-${24}$ & 15.11 & 3.03& 4501.2 & 3.17&  3.32 & -- & -- & --\\
F-${1}$ &14.81&  2.23&  487.1 & 2.34&  1.23 & -- & -- & --\\
F-${5}$ &--&1.22 &  77.1 &1.80 &2 & --& --&--\\
F-${12}$ & 13.66  &1.24 &  72.9&  1.77 & 1.25&--&--&--\\
F-${11}$ & -- &1.70 & 0.98 &  49.1&2&--&--&--\\
F-${9}$ &--&1.48 & 3.14 &  95.0&2&--&--&--\\
F-${17}$ &--&1.70  &0.90 &  45.0&2&--&--&--\\
F-${7}$ &16.23 & 7.03 &  96.3 & 1.14 & 2.79&--&--&--\\
F-${22}$ &11.67 & 0.39 & 102.8 & 2.52 & 1.20&--&--&--\\
F-${6}$ &--&1.57&  2.29 &  85.0&2&--&--&--\\
F-${34}$ &--&1.51&   1.30 & 42.2&2&--&--&--\\
F-${51}$ &--& 2.45&  0.49 & 138.8&2&--&--&--\\
F-${53}$ &--&1.95 & 0.91  & 81.5&2&--&--&--\\
F-${59}$ &--&1.48 & 1.97 &  59.5&2&--&--&--\\\hline
\end{tabular}
\end{center}
\label{tablestruct}
\end{table*}

\begin{table*}
\caption{Measured velocity dispersions $\sigma_{\rm obj}$ for the Fornax
UCDs without available HST imaging. IDs are from literature sources as in Table~\ref{table1}. $^*$This source designation
is from Richtler et al. (2008).}
\begin{center}
\begin{tabular}{|ll|rrrrrr|}
\hline
ID$_{\rm FLAMES}$ & ID$_{\rm literature}^*$ &RA [2000] & DEC [2000]&V$_0$ [mag] & M$_V$ [mag] & v$_{\rm rad,\sun}$ [km/s] &$\sigma_{\rm obj}$ [km/s] \\\hline
F-${3}$ & UCD27$_{\rm DW}$ &  3:38:10.4 & -35:24:06.2 &  19.7 & -11.7 & 1626 &  31.3 (1.5) \\
F-${18}$ & UCD44$_{\rm DW}$&  3:38:42.0 & -35:33:13.0 &  19.7 & -11.7 & 2024 &  19.1 (1.4) \\
F-${23}$ & UCD49$_{\rm DW}$&  3:39:20.5 & -35:19:14.2 &  19.7 & -11.7 & 1480 &  21.9 (1.4) \\
F-${2}$ & FCOS 2-2153 &  3:38:06.5 & -35:23:04.0 &    20.0 & -11.4 & 1426 &  18.7 (1.4) \\
F-${8}$ & FCOS 0-2066 &  3:38:23.2 & -35:20:00.7 &  20.1 & -11.3 & 1414 &  25.7 (1.4) \\
F-${40}$ & 92.099$^*$ &  3:37:52.5 & -35:28:57.9 &  20.7 & -10.7 & 1497 &  27.3 (1.4) \\
F-${60}$ & FCOS 2-2100 &  3:38:00.2 & -35:30:08.2 &  20.9 & -10.5 &  871 &  24.3 (1.9) \\
F-${64}$ & FCOS 1-2080 &  3:38:41.4 & -35:28:46.6 &    21.0 & -10.4 & 1728 &  24.7 (1.7) \\\hline
\end{tabular}
\label{table1a}
\end{center}
\end{table*}

\subsection{Mass modelling taking into account aperture effects}
To estimate the masses of the UCDs, we used
the mass modelling algorithm outlined in Hilker et al. (2007).
This includes a correction of the measured dispersion $\sigma_{\rm
obj}$ to the true global dispersion $\sigma_{\rm obj,cor}$, due to the
fact that our measurements miss the contributions from stars outside
the fibre aperture.
\vspace{0.15cm}

The mass modelling involved the following steps:
\vspace{0.1cm}

\begin{enumerate}
\item The observed, PSF-deconvolved luminosity profile from HST
  photometry was parameterized by the best-fitting density law. For
  most UCDs a satisfactory fit was achieved with a King or generalized
  King profile. Only UCD3 (ID$_{\rm FLAMES}$=F-19) required a
  two component King+Sersic function to be fitted well (see also
  Evstigneeva et al. 2007). The profile parameters for all 15 sources
  are shown in Table~\ref{tablestruct}.

\item The 2-dimensional surface density profile was deprojected by means of
Abel's integral equation into a 3-dimensional density profile.

\item The cumulated mass function $M(<r)$, the potential energy $\phi(r)$ and
the energy distribution function $f(E)$ were calculated from the 3-dimensional
density profile.

\item Finally, an N-body representation of the UCD was created by using the
deprojected density profile and the energy distribution function. For every 
model, 100.000 test particles were distributed and their $x$, $y$ and $z$ 
positions and corresponding $v_x, v_y$ and $v_z$ velocities were given as 
output.
\end{enumerate}

After generating the UCD model, the velocity dispersion as seen by an
observer was simulated. In doing so, the following steps were 
performed:

\begin{enumerate}
\item All test particles are convolved with a Gaussian whose full-width at
half-maximum (FWHM) corresponds to the observed seeing.

\item The fraction of the `light' (Gaussian) that falls into the fibre aperture
($1\farcs2$ for FLAMES) at the projected distance of the observed object 
(19 Mpc) is calculated.

\item These fractions are used as weighting factors for the velocities. All
weighted velocities that fall into the fibre region are then used to calculate
the `mimicked' observed velocity dispersion $\sigma_{\rm mod}$.
\end{enumerate}

Iteratively, the total `true' mass of the modelled object, $M_{\rm true}$, 
that corresponds to the observed velocity dispersion, $\sigma_{\rm obs}$ is
then determined by scaling a first `guess' mass, $M_{\rm guess}$, via the
formula $M_{\rm true} = M_{\rm guess}\cdot (\sigma_{\rm obs}/\sigma_{\rm 
mod})^2$.

\label{modelling}

The masses, mass-to-light ratios, global and central
velocity dispersions and the main model parameters derived in this way are listed in
Table~\ref{table1}. Note that the three brightest UCDs in our study
were also observed by Hilker et al. (2007) with UVES. The M/L ratios derived for
them in that study are indicated in Table~\ref{table1}. The error ranges
of the three estimates overlap, and the
average ratio between our M/L values and those from Hilker et al.
(2007) is 0.90 $\pm$ 0.16, consistent with unity.

\section{Results}
\label{results}

\begin{figure}[h!]
\begin{center}
  \epsfig{figure=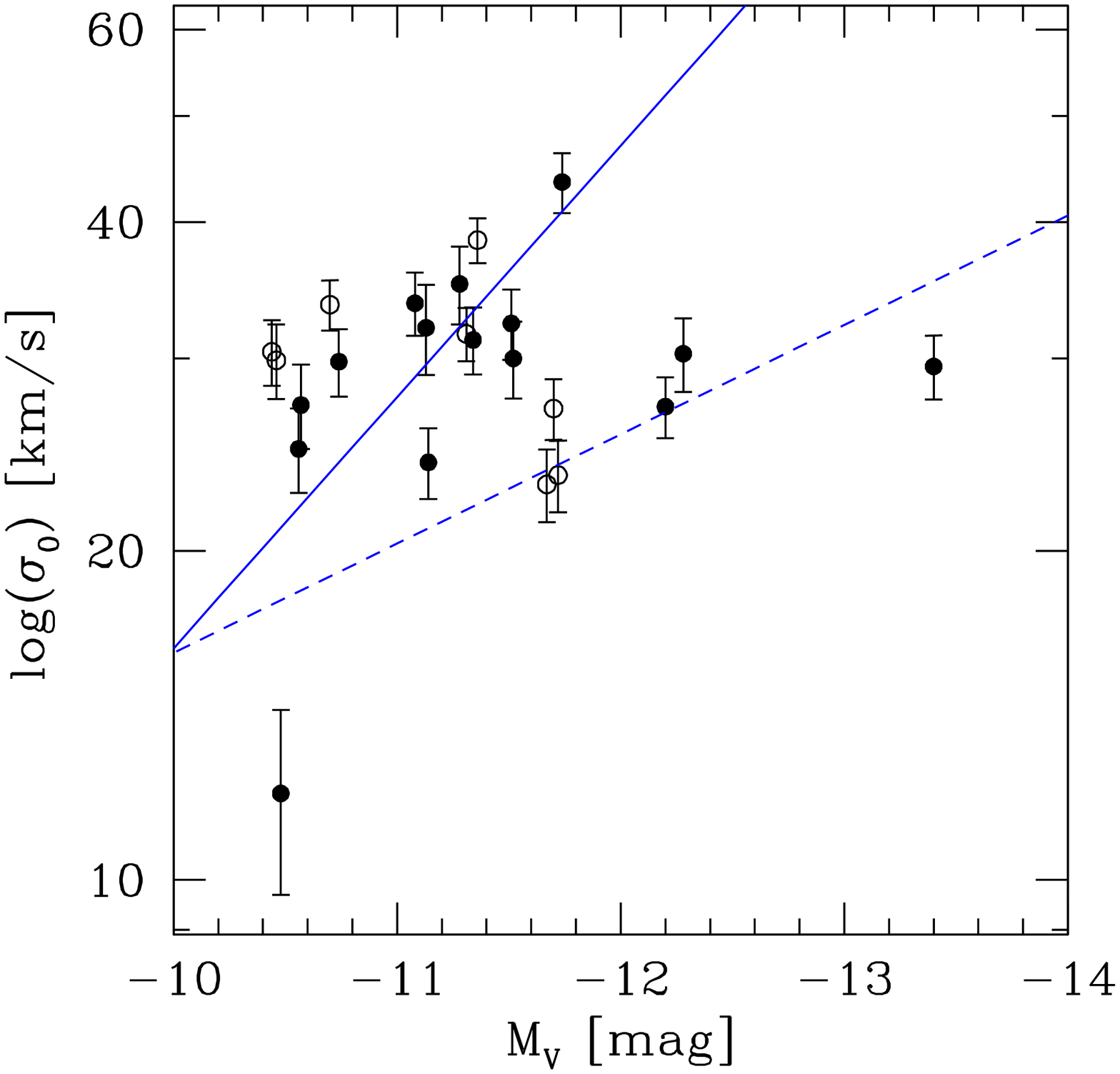,width=8.6cm}

  \caption{Absolute magnitude of UCDs plotted against their central velocity
    dispersion $\sigma_0$. The solid line is the extrapolation of the
    relation defined by Galactic GCs (McLaughlin \& van der Marel
    2005), the dashed line is the extrapolation of the Faber-Jackson
    relation for luminous elliptical galaxies (Faber \& Jackson 1976),
    which also fits the compact elliptical M32. Open symbols
    indicate objects without HST imaging, for which the average
    correction factor between global $\sigma$ and $\sigma_0$ was
    assumed as derived from modelling of the 15 objects with HST
    imaging.}
\label{MV_sigma}
\end{center}
\end{figure}

With the total mass derived from the dynamical modelling, we
calculated the optical mass-to-light ratio M/L$_V$, using the V-band
photometry from the wide-field imaging data presented in Hilker et al.
(2003) and Mieske et al.~(2006, 2007a).  Based upon the modelling
algorithm of Hilker et al. (2007), we also calculated the central
velocity dispersion $\sigma_0$ for all sources with HST imaging
available. The ratio $\frac{\sigma_0}{\sigma}$ was 1.23 on average
with a scatter of 0.07.  For those eight sources with reliable $\sigma$
measurements but without HST data (Table~\ref{table1a}), we assumed
an average correction factor of 1.23 to include them in an analysis
of their locus in the $M_V - \sigma_0$ plane. This plot is shown in
Fig.~\ref{MV_sigma}.  

In the $M_V - \sigma_0$ plane, the relation for globular clusters and
the extrapolation of the Faber-Jackson relation for luminous elliptical
galaxies (Faber \& Jackson 1976) intersect at about $M_V\simeq -10$
mag. In the luminosity regime of our sample of Fornax UCDs, both
extrapolations bifurcate. We can therefore roughly subdivide our sample into
objects closer to the extrapolation of GC relation, and objects closer
to the extrapolation of the Faber-Jackson relation (which also happens to match
the compact elliptical galaxy M32; Evstigneeva et al. 2007).  The
three brightest UCDs ($M_V<-12$ mag) are clearly more consistent with
the Faber-Jackson relation while fainter UCDs are preferentially closer to
the GC relation. We find that the projected clustercentric distance of
objects that are more consistent with the Faber-Jackson relation is
$\sim 60 \pm 30$\% larger than that of sources more consistent with
the GC relation. These findings are consistent with the
hypothesis that our sample consists of both objects associated to the
globular cluster system of NGC 1399, and objects with more complex
dynamical formation history, being associated more to the overall
cluster potential.

\begin{figure}[]
\begin{center}
  \epsfig{figure=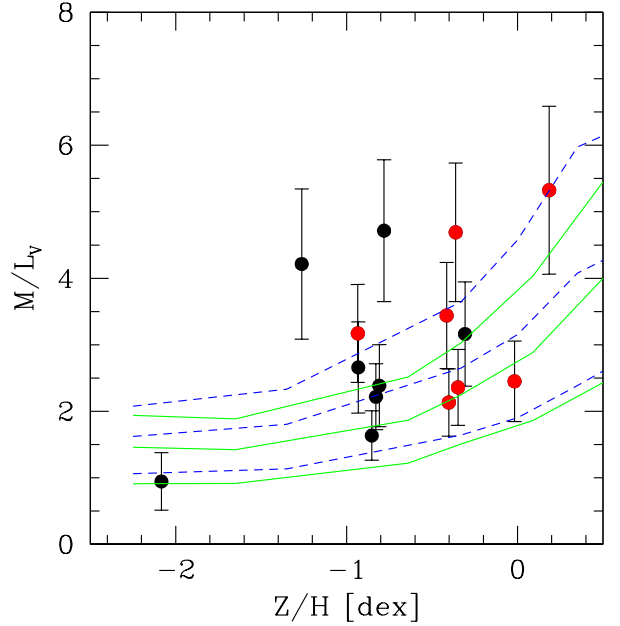,width=8.6cm}
  \caption{Z/H of the UCDs plotted against their M/L ratios. Solid (green) lines indicate stellar population models from Bruzual \& Charlot~(2003) for ages (from bottom to top) of 5, 9, and 13 Gyrs, assuming a Chabrier IMF. Blue (dashed) curves are from Maraston et al.~(2005) for the same age ranges, assuming a Kroupa IMF. The Z/H values are from Mieske et al.~(2006), assuming a solar [$\alpha$/Fe] abundance, as suggested in Mieske et al.~(2007b). Red data points indicate sources for
which spectroscopic [Fe/H] estimates are available. For the remaining sources,
[Fe/H] is estimated from their (V-I) colour, using the transformation of Kissler-Patig et al.~(1998) (see also Mieske et al.~2006).}
\label{VI_ML}
\end{center}
\end{figure}

In Fig.~\ref{VI_ML} we plot metallicity Z/H against M/L ratio. The
metallicity Z/H is derived directly for some sources from previous
spectroscopy (Mieske et al. 2006), for others derived from their (V-I)
colour (Mieske et al. 2007a), using the calibration of Kissler-Patig
et al. (1998). This calibration was shown to be accurate to within
0.1-0.2 dex (Mieske et al. 2006) for old stellar populations with
$\rm [Fe/H]\gtrsim-$1.0 dex. In the plot we indicate SSP predictions from
Bruzual \& Charlot (2003) and Maraston et al. (2005) for M/L ratios of
populations with solar [$\alpha$/Fe] abundances, with ages between 5
and 13 Gyrs. The former models assume a Chabrier IMF (Chabrier 2003),
while the latter models assume a Kroupa IMF (Kroupa 2001). We note
that both IMFs are very similar and do not account for the
difference in predicted M/L ratios at fixed age and metallicity (see
also Dabringhausen et al.  2008).  Rather, it is the choice of
different stellar evolutionary codes which leads to the 20\%
differences between Bruzual \& Charlot and Maraston M/L predictions.
Most of the M/L data points are consistent to within their errors with
the theoretical predictions assuming a canonical IMF, with three sources
showing somewhat elevated M/Ls.

\begin{figure*}[]
\begin{center}
  \epsfig{figure=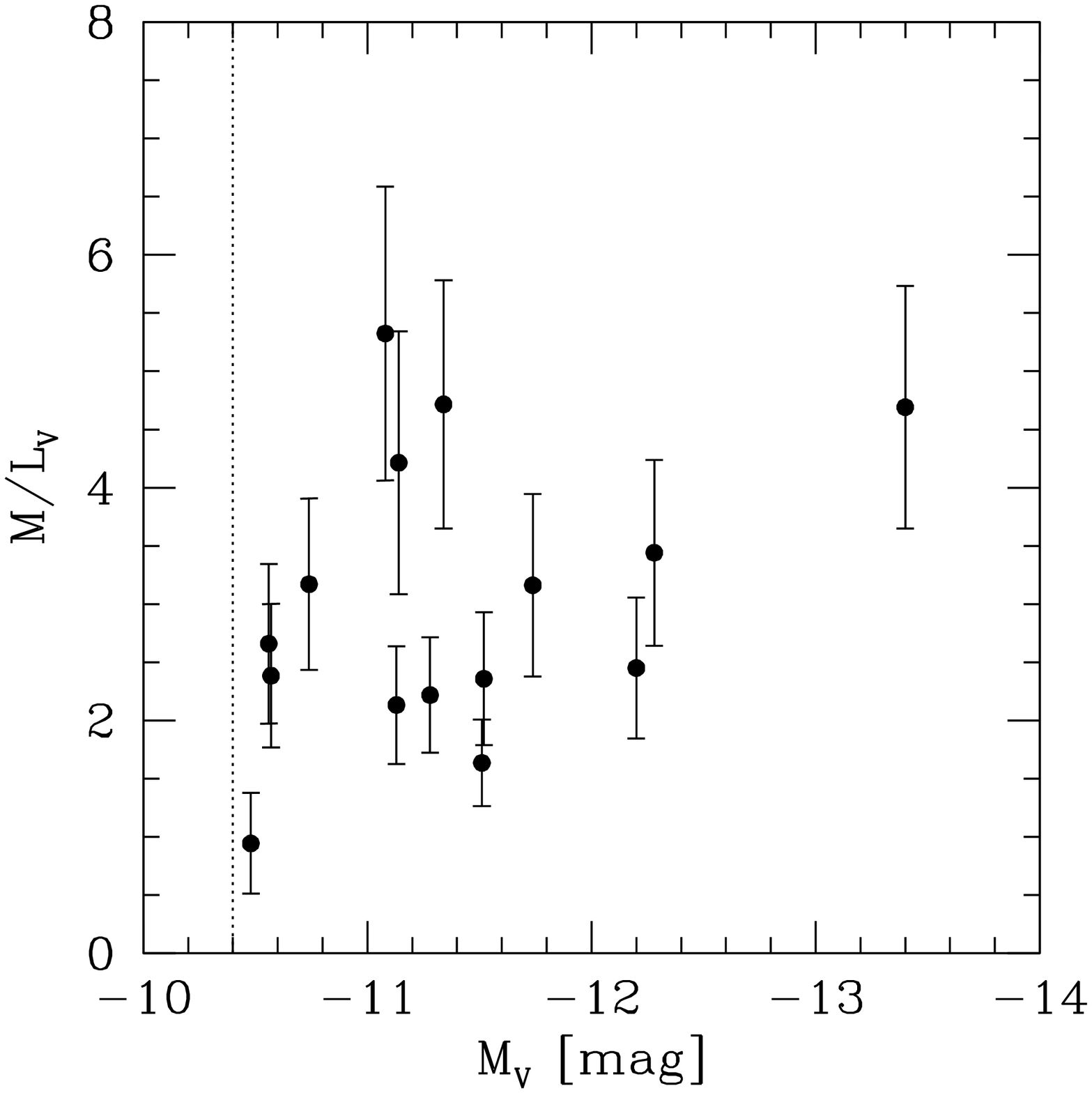,width=8.6cm}
  \epsfig{figure=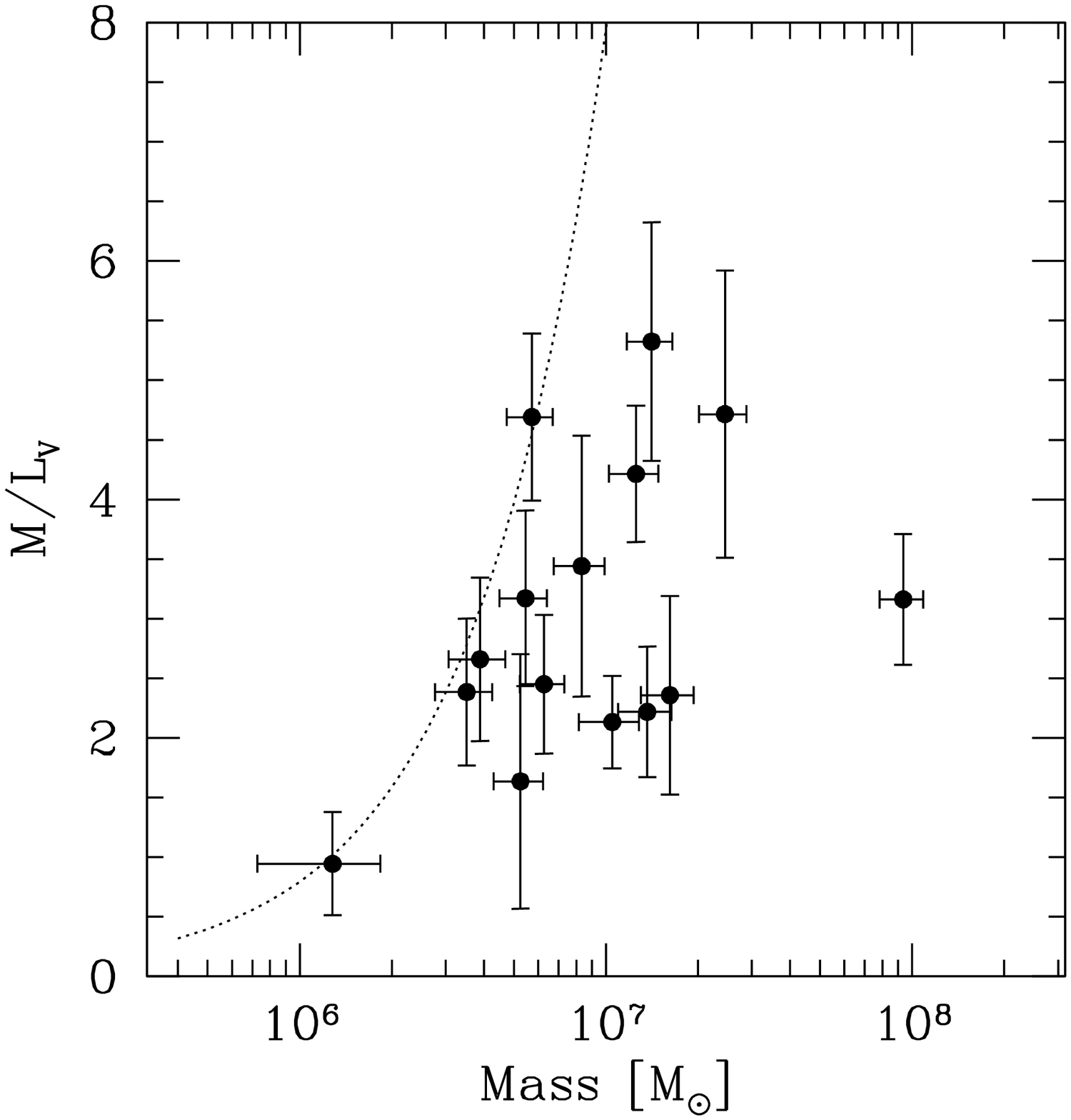,width=8.6cm}
  \epsfig{figure=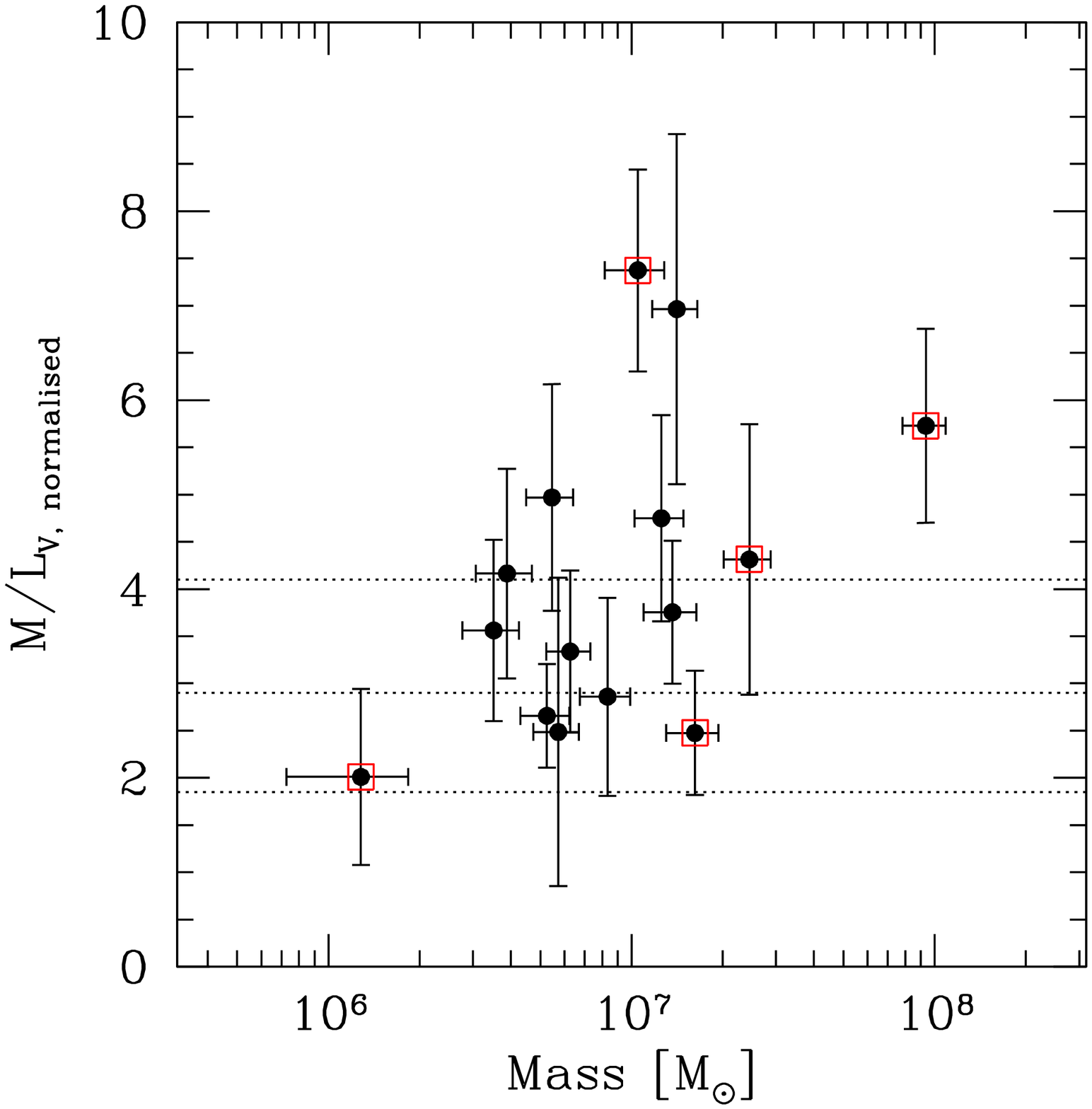,width=8.6cm}
  \epsfig{figure=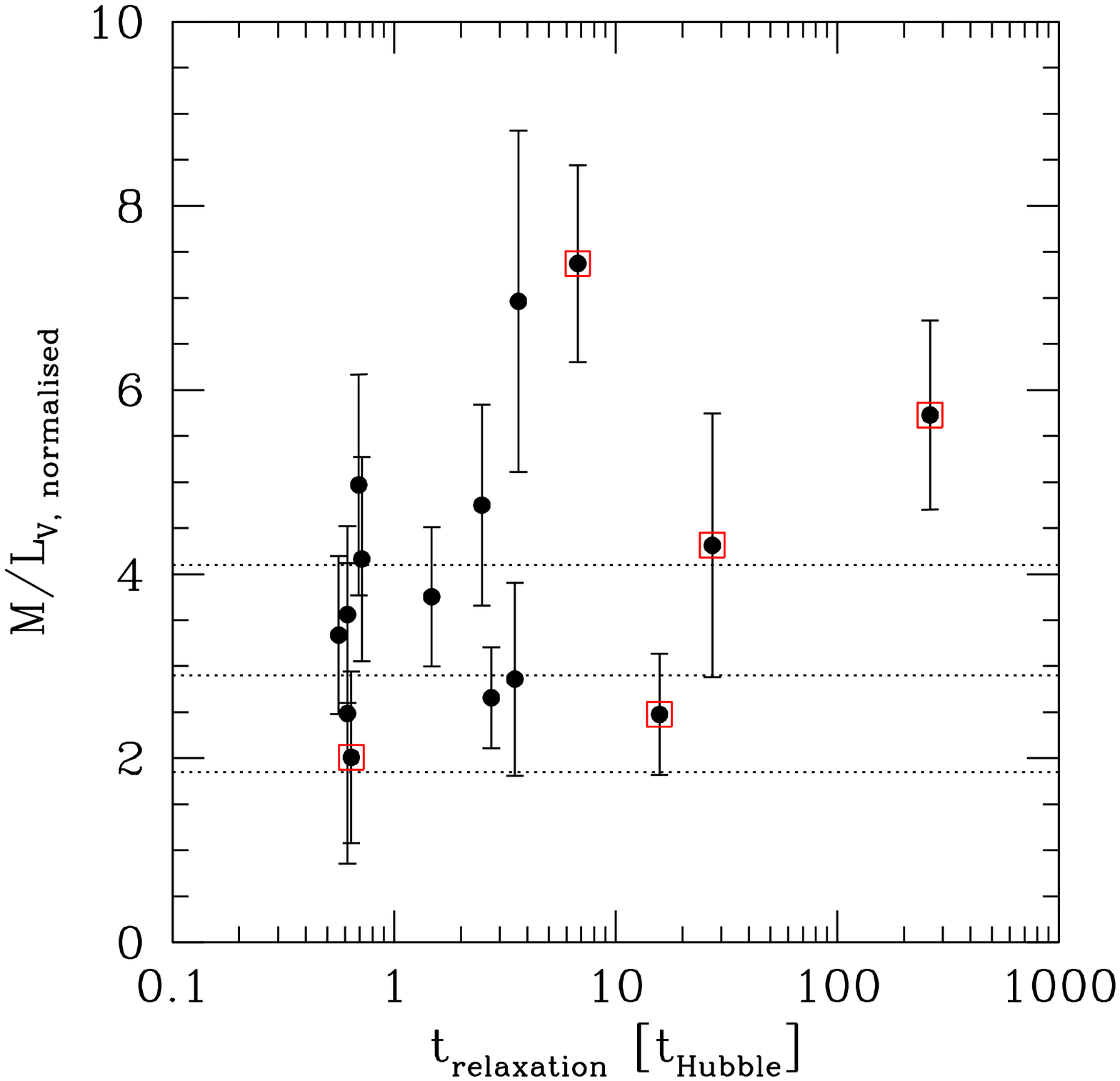,width=8.6cm}
  
  \caption{{\bf Top Left:} Absolute magnitude of UCDs plotted against their M/L ratio
in the V-band. The dotted line indicates the faint magnitude limit of our survey. {\bf Top Right:} Masses of the UCDs from the left plot, plotted against their M/L ratio. The dotted line indicates the mass dependent upper limit of our M/L sensitivity, caused by our faint magnitude limit of $M_V=-10.4$ mag. There is a 3$\sigma$ correlation between the shown data points. {\bf Bottom left:} Plot analogous to the top right panel. Now, the M/L ratio measurements have been
normalised to the same (solar) metallicity (see text). For this, we assume the mean of the model predictions from Bruzual \& Charlot (2003) and Maraston (2005) in Fig.~\ref{VI_ML}. The horizontal lines indicate the M/L ratios from the model predictions for 13, 9, and 5 Gyrs (top to bottom). There is only a 1.5$\sigma$ correlation between the data points.  Sources marked by large squares are those closer to the Faber Jackson relation in Fig.~\ref{MV_sigma}. {\bf Bottom right:} As in the plot on the left, but now the  x-axis is relaxation time.}

\label{MV_mass_ML}
\end{center}
\end{figure*}

In the two top panels of Fig.~\ref{MV_mass_ML} we plot $M_V$ and mass
vs. the M/L ratio for the 15 sources with HST imaging. We indicate the
faint magnitude limit of our survey, which translates into a mass
dependent M/L sensitivity limit.  To test whether the rise of M/L with
mass generally observed in the regime $10^5<M/M_{\rm \sun}<10^8$ can be
traced by our data, we fit a linear relation to the distribution of
mass vs. M/L ratio. We find a slope different from 0 at the 2.8$\sigma$
level. The significance of the slope was calculated by random
resampling of the data points around the fitted relation. For this
re-sampling, the scatter of the data points around the relation was
used, given that it was about 25\% larger than the average error of
the data points.

There are two caveats regarding the interpretation of this
$\sim$3$\sigma$ slope.  The first caveat is the mass dependent M/L
sensitivity limit.  Assuming a random distribution of M/L ratio with
$M_V$, a fixed magnitude limit will artificially create a slope in the
mass $-$ M/L plane, due to the lack of sources at low masses and high
M/L ratio.  The second caveat is that a relation between mass and M/L
is naturally produced if there is a relation between mass and
metallicity (see e.g.  Mieske et al. 2004 and 2006). This is because
optical M/L ratios increase towards higher metallicites (see
Fig.~\ref{VI_ML}). We correct for this effect by normalising our
M/L ratios to solar metallicity (see also Mieske \& Kroupa 2008, and
Dabringhausen, Hilker \& Kroupa 2008).  To this end, we fit a relation

\begin{equation}
M/L_{\rm theo}=a+b*exp(c*{\rm [Fe/H]})
\end{equation}

\noindent to the M/L predictions for a
13 Gyr population from each of the two model sets of Bruzual \&
Charlot (2003) and Maraston (2005) (see Fig.~\ref{VI_ML}). We 
define the mean of the two fits as the reference relation:
$M/L_{\rm theo}=0.5 \times (M/L_{\rm theo,BC03}+M/L_{\rm theo,M05})$. See Mieske
\& Kroupa (2008) for the details of these fits. We then normalise our
M/L values in the following way:

\begin{equation}
(M/L)_{\rm normalised}=\frac{(M/L)}{(M/L)_{\rm theo}}*(M/L)_{\rm theo,0}
\end{equation}

\noindent $(M/L)_{\rm theo,0}$ is the theoretical prediction for [Fe/H]=0.

The normalised values are plotted vs. mass in the bottom left panel of
Fig.~\ref{MV_mass_ML}. Their errors include the difference between the
model predictions and an assumed uncertainty of 0.3 dex in [Fe/H],
although note that these contributions are small compared to the
uncertainty of the M/L measurement itself. The significance of the
correlation decreases to 1.5$\sigma$ (and 1.2$\sigma$ when excluding
the lowest mass data point). Our data are hence consistent with a
non-correlation between mass and M/L ratio in the mass range $3\times
10^6 < M < 10^8$ M$_{\rm \sun}$, especially when taking into account
the mass dependent selection limit.  We find no extremely high M/L
ratios as in Virgo (Ha\c{s}egan et al. 2005).

In the bottom right panel of Fig.~\ref{MV_mass_ML} we plot the
normalised M/L ratio vs. half-mass relaxation time $t_{\rm relax}$ (see also
Table~\ref{table1}).  The relaxation time is a more direct measure of
the state of dynamical evolution than the mass. We use the following
equation from Dabringhausen, Hilker \& Kroupa (2008, equation 6 in
their paper) to calculate $t_{\rm relax}$ in units of Myrs:
\begin{equation}
t_{\rm relax}=\frac{0.234}{log(M)}*\sqrt{M*r_{\rm \rm eff}^3/0.0045}
\end{equation}

\noindent This formula is based on Spitzer \&
Hart (1971) and Spitzer (1987). The mass $M$ is given in solar masses
and $r_{\rm \rm eff}$ in pc. We find only a very marginal trend
(2.1$\sigma$) of increasing M/L ratio with increasing $t_{\rm relax}$.
A larger sample over a broader mass range is required to quantify a
trend of M/L with mass or relaxation time (Sect.~\ref{MLdiscussion}).

\begin{figure}[h!]
\begin{center}
  \epsfig{figure=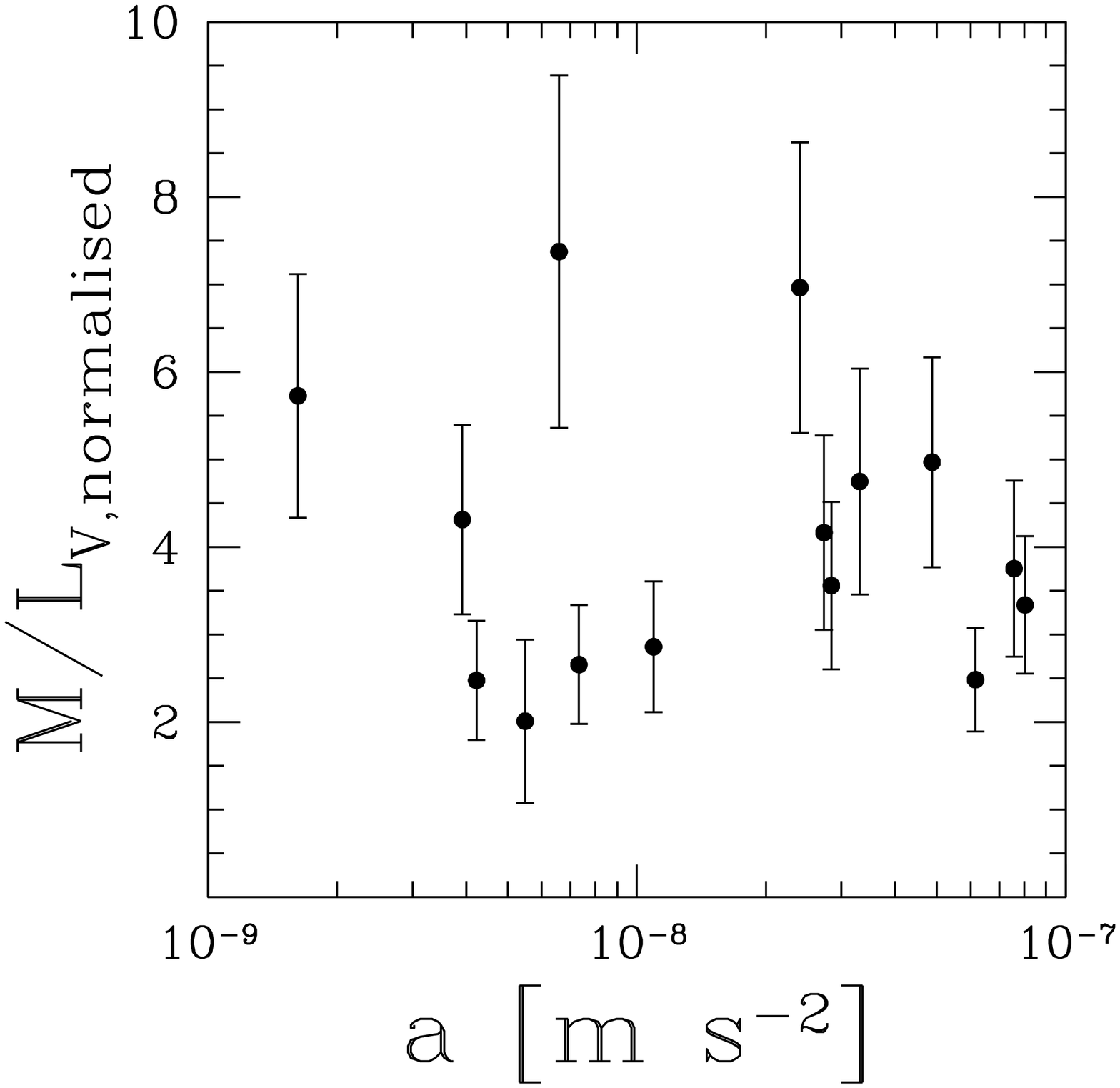,width=4.3cm}
  \epsfig{figure=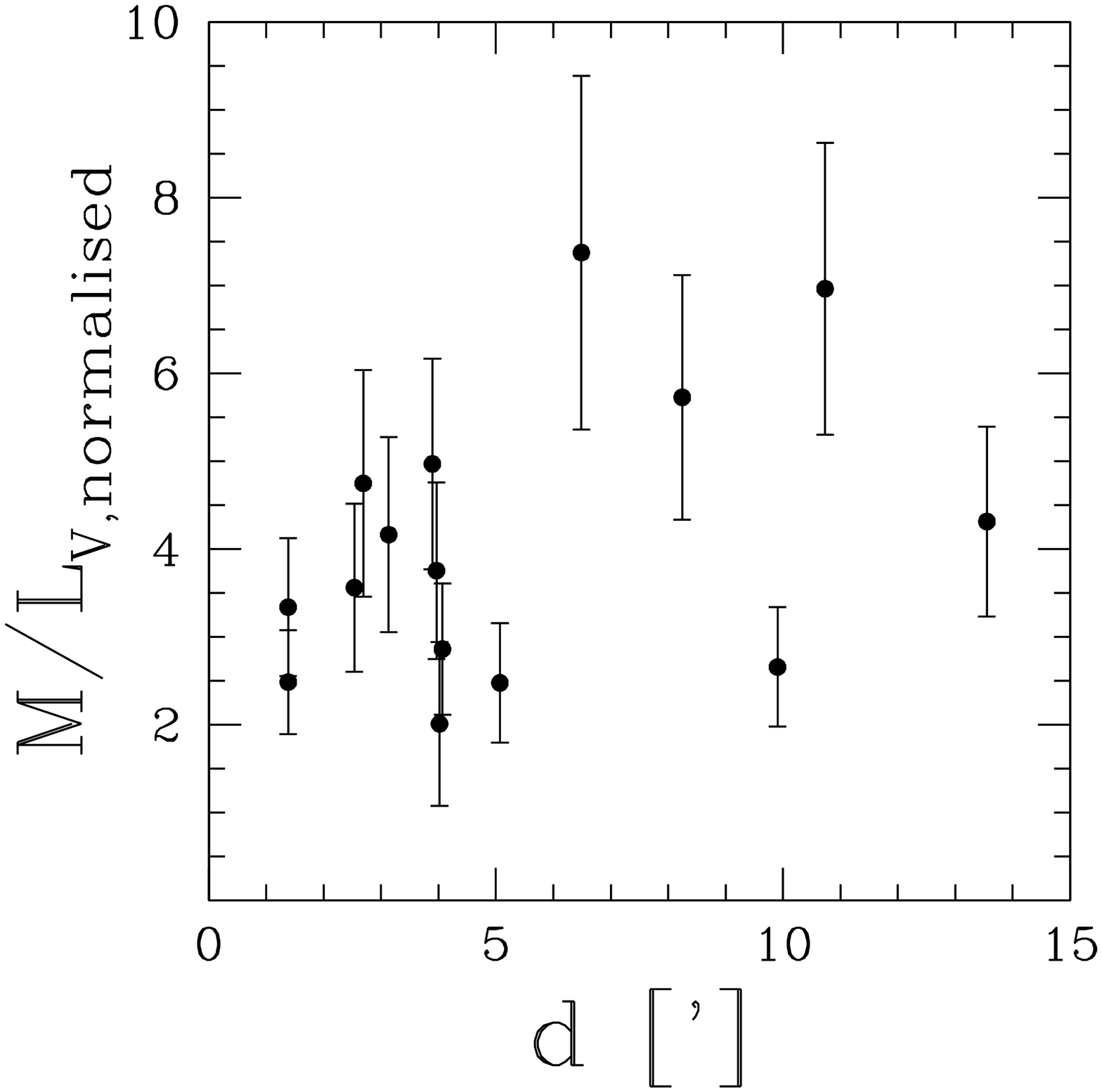,width=4.3cm}
  \caption{{\bf Left panel}: Gravitational acceleration, $a=\frac{G \times M}{r_h^2}$, plotted against normalised M/L ratio. None of
    the objects are in the low acceleration regime $a\lesssim
    1.2*10^{-10}$m s$^{-2}$ where MOND has been postulated to hold.
    {\bf Right panel:} Projected distance to NGC 1399 plotted against
     normalised M/L ratio.}

\label{ML_a}
\end{center}
\end{figure}

In Fig.~\ref{ML_a} we plot M/L ratio vs. the acceleration parameter
$\rm a=\frac{G \times M}{r_h^2}$ to illustrate that the internal
dynamics of the compact objects are far from the MOND (Milgrom 1983)
regime of weak acceleration ($\rm a_0 \sim 1.2 \time 10^{-10}$ m
s$^{-2}$).  Fig.~\ref{ML_a} also shows the M/L ratio as a function of
projected distance d to NGC 1399.  Table~\ref{table1} lists both
  the acceleration parameter a and the projected distance d. In the
case that tidal heating efficiently increases the internal velocity
dispersion of sources with small apocentric radii (Fellhauer \& Kroupa
2006), one may expect a trend of increasing M/L with decreasing
radius. No such trend is seen.

\section{Discussion}
\label{discussion}

\subsection{M/L ratio measurements over a range of environments}
\label{MLdiscussion}
In order to further quantify the dependence of M/L ratio on mass, and
to investigate M/L variation with environment, we combine our data for
Fornax UCDs with M/L ratio measurements of other compact stellar
systems from literature studies, see Fig.~\ref{mass_ML_all}. This
covers the regime of low-mass Milky Way globular clusters (M$\gtrsim 5
\times 10^4$M$_{\rm \sun}$) up to the most massive UCDs
(M$\sim$$10^8$M$_{\rm \sun}$).

\begin{figure}[]
\begin{center}
\epsfig{figure=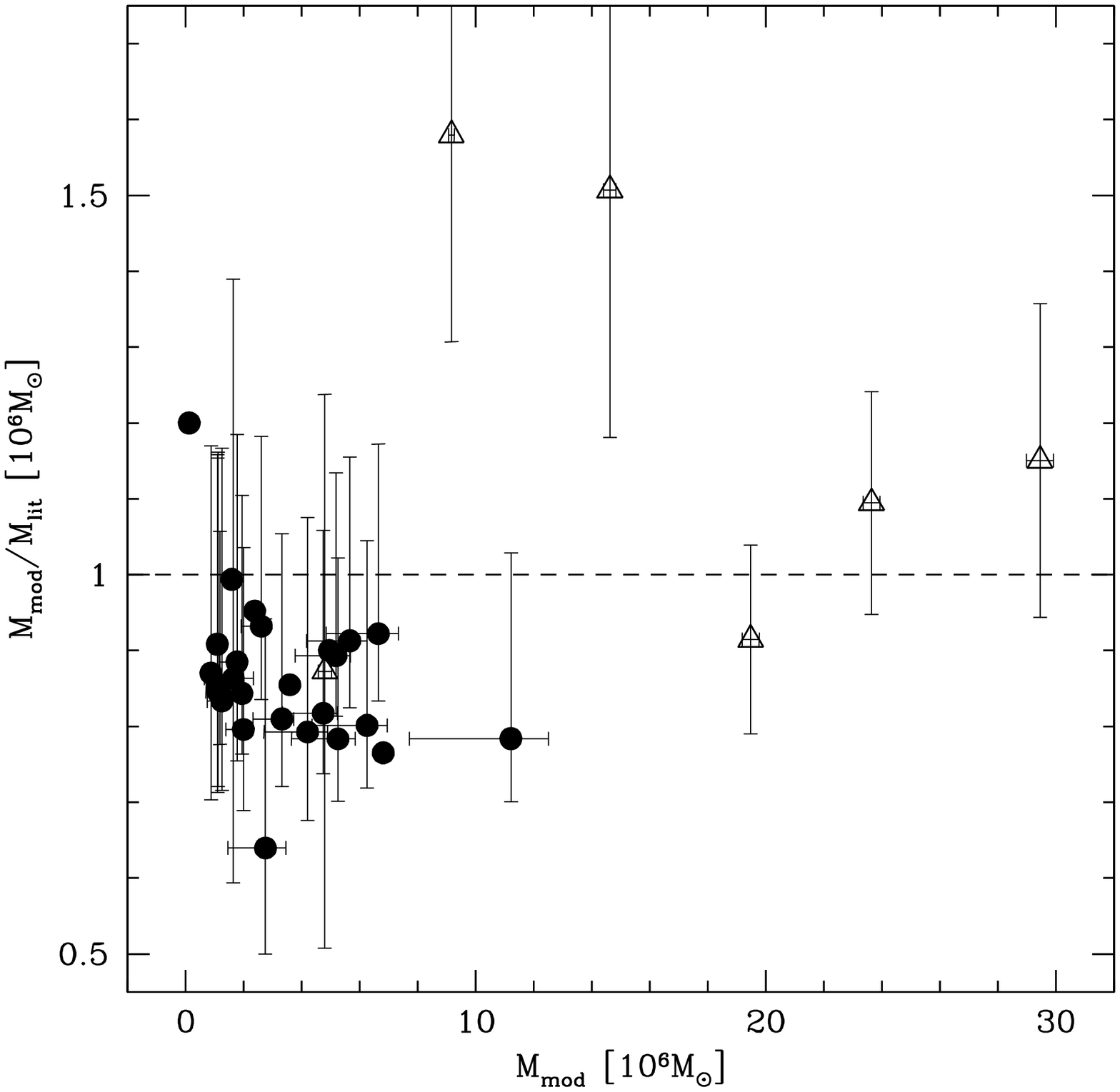,width=8.6cm}
\caption{The effect of proper aperture corrections on dynamical mass estimates. 
The ratio between modelled mass and dynamical mass as given in the literature 
is plotted vs. modelled mass. The masses of Cen\,A GCs from Rejkuba
et al. (2007) (solid circles) were previously overestimated by up to 30\%, 
whereas the masses of DGTOs in Virgo from Ha\c{s}egan et al. (2005) (open 
triangles) were mostly underestimated. The errorbars reflect the uncertainties
given for the literature values.}
\label{mass_mod_diff}
\end{center}
\end{figure}
\begin{figure}[]
\begin{center}
\epsfig{figure=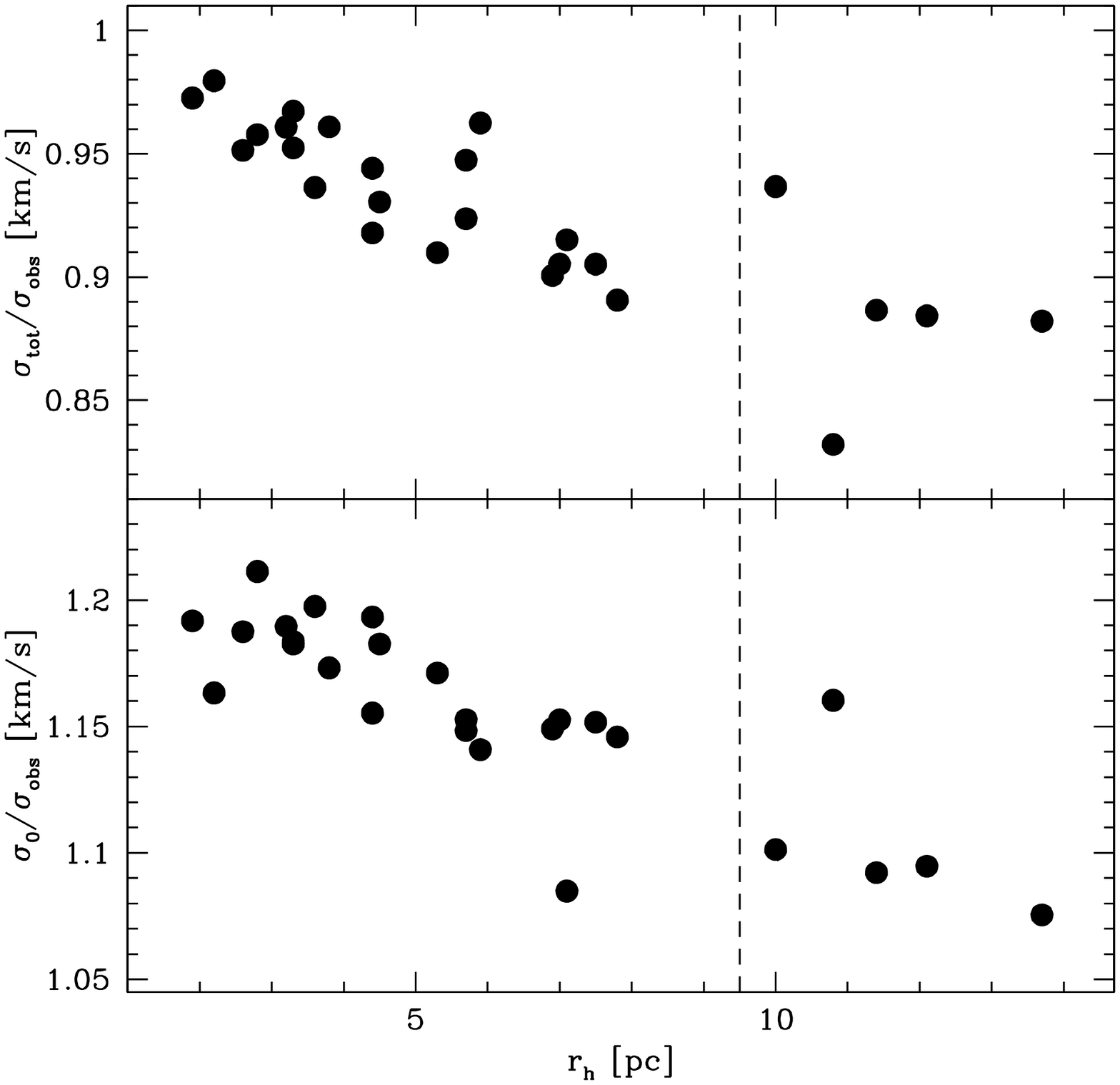,width=8.6cm}
\caption{Corrections from observed to total and central velocity dispersions
for Cen\,A GCs based on our mass modelling. The upper panel shows the ratio 
between total and observed velocity dispersion, while the lower panel shows the ratio 
between central and observed velocity dispersion. The dashed vertical line 
indicates half the width of the slit aperture.
}
\label{sig_cena}
\end{center}
\end{figure}

\subsubsection{Revision of literature mass estimates}
\label{revision}
\begin{figure*}[]
\begin{center}
  \epsfig{figure=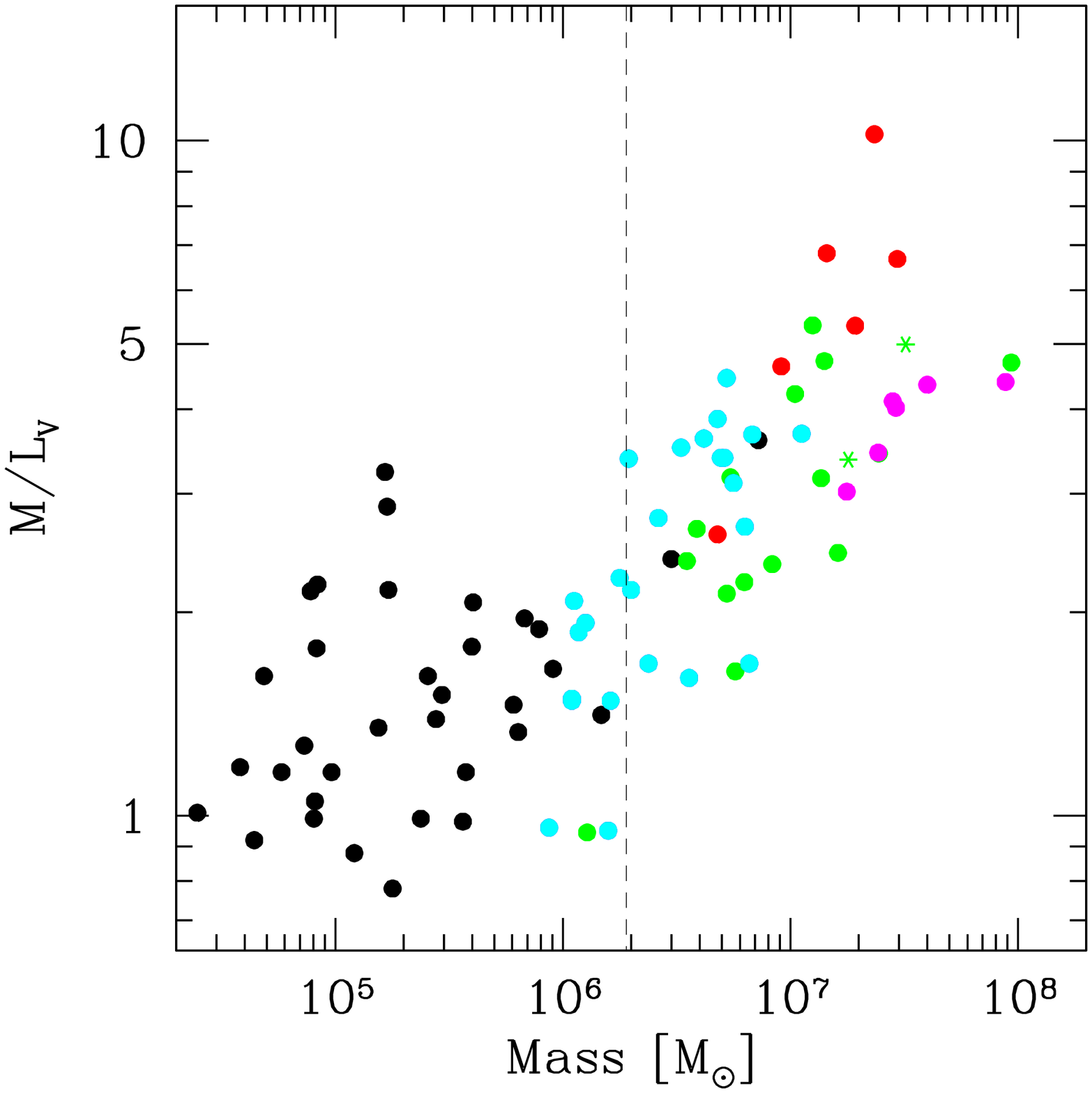,width=8.6cm}
  \epsfig{figure=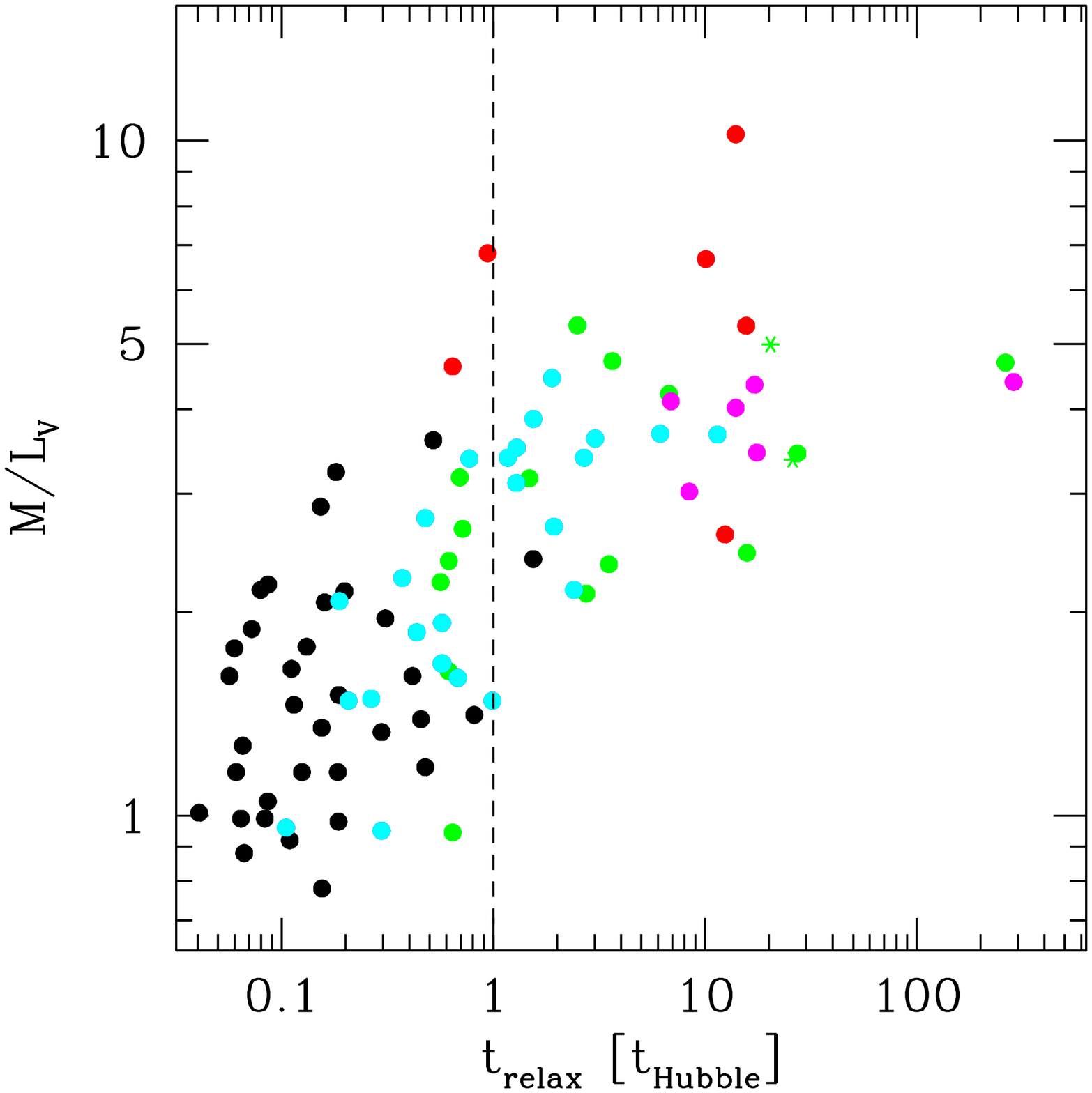,width=8.6cm}
  
  \epsfig{figure=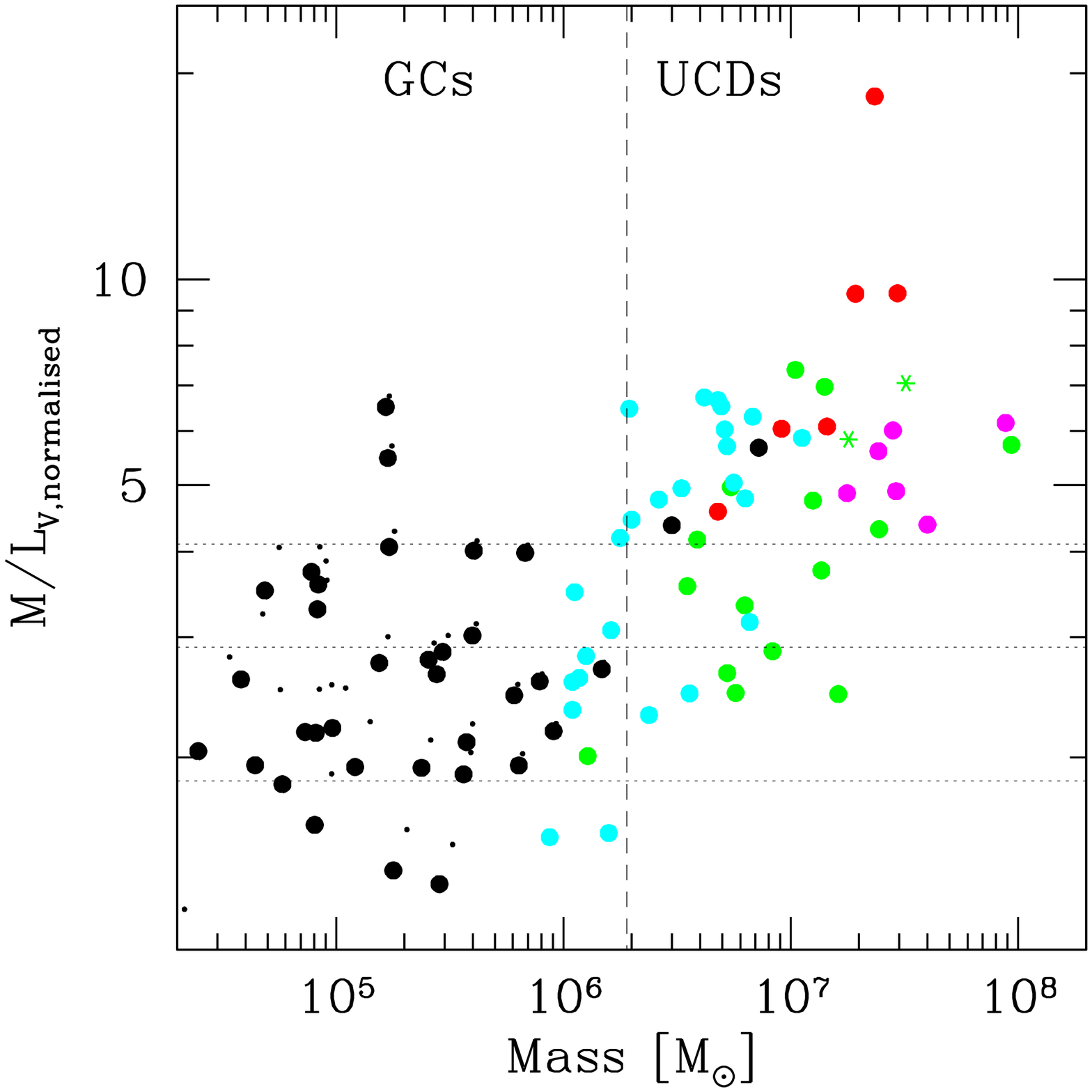,width=8.6cm}
  \epsfig{figure=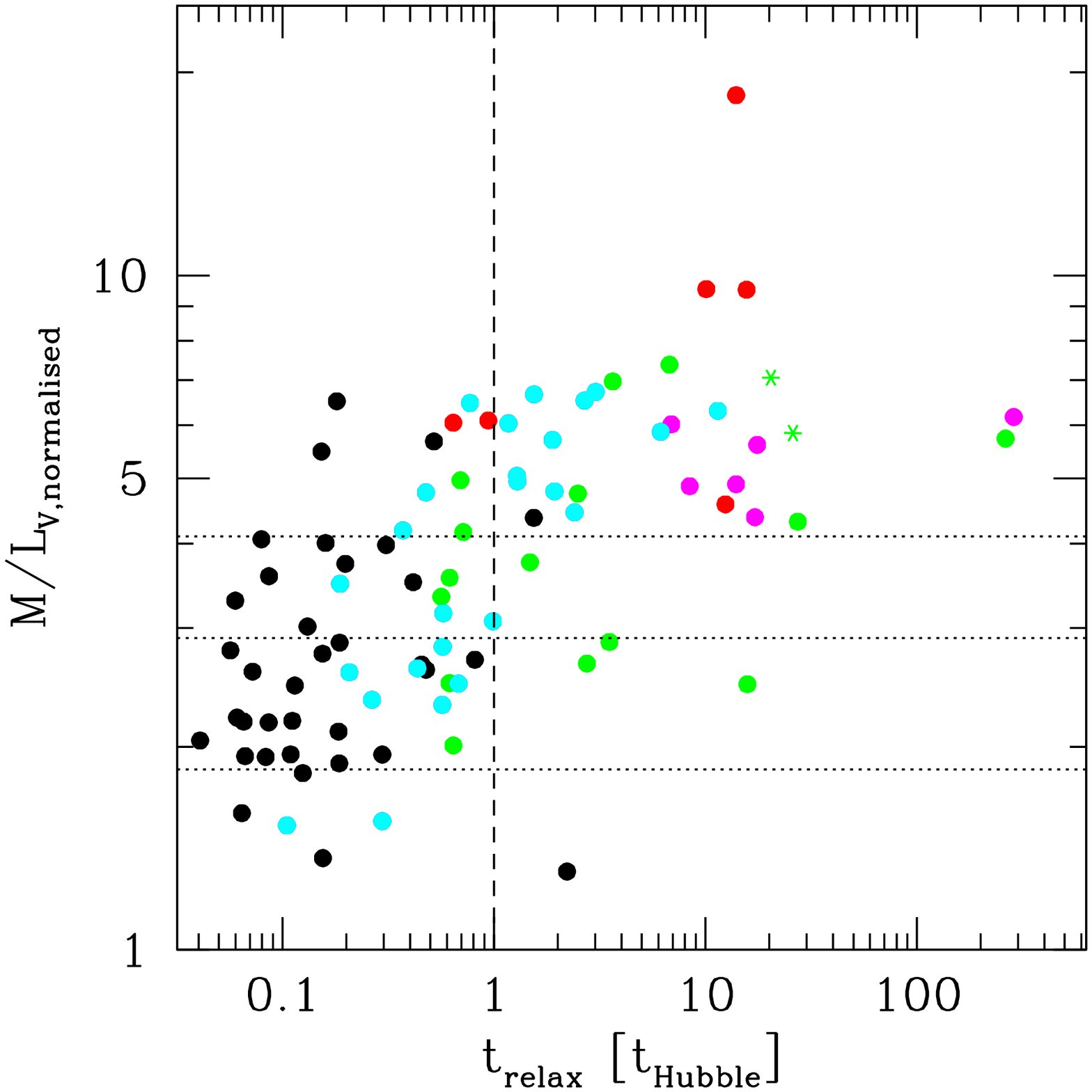,width=8.6cm}
  \caption{{\bf Top left panel:} Mass vs. M/L for the isolated compact objects of this paper (green dots) plus literature values of compact objects in CenA (Rejkuba et al. 2007, cyan), Virgo (Ha\c{s}egan et al. 2005, red; Evstigneeva et al. 2007, magenta), Fornax (Hilker et al. 2007, green asterisks), and Milky Way globular clusters (McLaughlin \& van der Marel 2005, black dots).  The vertical dashed line indicates the approximate mass where the relaxation time is equal to one Hubble time (see also the right panels). {\bf Top right panel}: Like the left panel, but plotting relaxation time instead of mass on the x-axis.  {\bf Bottom left panel:} Like the top left panel, but here all M/L ratio estimates have been normalised to solar metallicity (see text). The horizontal dotted lines indicate the M/L ratios expected for single stellar populations of age 13, 9, and 5 Gyrs (from top to bottom) based on Bruzual \& Charlot (2003) and Maraston et al. (2005). The small (black) dots indicate the present-day M/L ratios of the Galactic GCs if they would not have undergone dynamical evolution (see text). {\bf Bottom right panel:} Like the left panel, but plotting relaxation time instead of mass on the x-axis. }
\label{mass_ML_all}
\end{center}
\end{figure*}

In Fig.~\ref{mass_ML_all} we use revised dynamical mass estimates of
the Cen\,A globular clusters from Rejkuba et al. (2007) and the DGTOs
from Ha\c{s}egan et al.  (2005). Those estimates are obtained by applying
the same mass modelling and aperture simulations as presented in
Sect.~\ref{modelling} and in Hilker et al. (2007). To this end, we
assumed King models as representations of the light profiles with the
projected half-light radii and King concentrations as given in the two
papers. We include new measurements from five additional Cen A
globular clusters (Rejkuba et al. 2007), using structural parameters
provided by M. Gomez from observations with IMACS on Magellan (private
communication).

It turned out that the modelled masses (and thus mass-to-light ratios)
of the Cen\,A globular clusters are up to 30\% lower than estimated
from the virial mass estimator, whereas most of the modelled DGTO
masses are in general higher (up to 60\%).  Figure~\ref{mass_mod_diff}
compares our modelled masses with the dynamical estimates given in the
literature. The reason for the discrepancies is the different treatment of the
aperture corrections that have to be applied to the observed velocity
dispersions.  

In case of the Cen\,A data the authors estimated the
aperture corrections to be a few percent, but they preferred to assign
these corrections only to the total error budget, and instead used
directly the observed ($\sigma_{\rm \rm obs}$) values with the virial
estimator to derive masses. However, at the distance of Cen\,A, 3.84
Mpc (Rejkuba 2004), the ratio between the projected half-light
diameter (4-48 pc) and the slit width ($1\arcsec \simeq 19$ pc) is
such that aperture corrections can not be neglected. A large fraction
of the light of the most extended massive clusters lies outside the
slit area which prohibits the measurement of a global velocity
dispersion.  The ratios $\sigma_{\rm \rm tot}/\sigma_{\rm \rm obs}$ and
$\sigma_0/\sigma_{\rm \rm obs}$ as function of projected half-light radius
are shown in Fig.~\ref{sig_cena}.

In case of the DGTO data, the authors derived the dynamical mass from
the King mass estimator (e.g., Dubath \& Grillmair 1997), for which
the core radius and the central velocity dispersion have to be known.
Ha\c{s}egan et al.  (2005) corrected the measured velocity dispersion to
the central one by ``scaling upward to account for the blurring of the
actual velocity dispersion profiles within the ESI slit'' for which the
knowledge of the intrinsic light profile from HST imaging has to be
known.  The Virgo cluster is far enough away, 16.1 Mpc (Tonry et al.
2001), that most of the light of the DGTOs, which have half-light
diameter between 7 and 58 pc, falls into the slit (with width of
$0\farcs75 \simeq 58$ pc). One tends to measure the global velocity
dispersion for most of the DGTOs. Thus, the correction to {\it
  central} velocity dispersions should be larger for {\it smaller}
objects, since for those the measured dispersion is closest to the
global one. This, however, is opposite to the trend of the corrections
applied by Ha\c{s}egan et al. (2005) (see their Tables 4 and 5), such that
on average the masses derived by Ha\c{s}egan et al. (2005) have to be
corrected upwards (see Fig.~\ref{mass_mod_diff}). We will take our
modelling results for the further discussions. The revised model
masses will also be included in an upcoming paper (Ha\c{s}egan et al.
2008, in preparation), along with M/L measurements of newly
discovered Virgo DGTOs.

\subsubsection{A trend of M/L with mass and relaxation time}
In Fig.~\ref{mass_ML_all}, we plot both, mass and relaxation time, vs.
direct and normalised M/L ratios for our compiled sample of compact
stellar systems. In Table~\ref{table2}, the M/L ratios, metallicities
and sizes for these objects are shown.  When fitting a linear relation
to relaxation time as a function of mass, we find that relaxation time
equal to 1 Hubble time is reached at a mass of about 2$\times
10^6$M$_{\rm \sun}$. In the following we adopt this mass as an
approximate limit between globular clusters and UCDs (see also Mieske
\& Kroupa 2008).

Fig.~\ref{mass_ML_all} shows a clear rise of the M/L ratio for masses
above $\simeq$$2\times10^6$ M$_{\rm \sun}$. When correcting the M/L
ratio measurements for their metallicity dependence, this is still
clearly visible (see also Mieske \& Kroupa 2008; Dabringhausen, Hilker
\& Kroupa 2008), formally significant at the 8 $\sigma$ limit. The M/L
ratio distribution of objects below and above the $2\times10^6$
M$_{\rm \sun}$ limit stems from the same parent distribution with only
the 2.7$\times 10^{-11}$ probability, according to a KS test.  The
average normalised M/L ratio for globular clusters is 2.70 $\pm$ 0.17,
while it is 5.44 $\pm$ 0.37 for UCDs. The rise in M/L corresponds to a
40\% increase of normalised M/L ratio per mass decade: $\frac{{\rm d}
  log(M/L))}{{\rm d} log(M)}=0.147 \pm 0.019$ (applying a 3$\sigma$
clipping).  A separation between UCDs and GCs at $t_{\rm relax}=$1
t$_{\rm Hubble}$ leaves the mean M/L ratios of GCs and UCDs unchanged
with respect to the mass cut at $2\times10^6$ M$_{\rm \sun}$. For the
separation at $t_{\rm relax}=$1 t$_{\rm Hubble}$, the M/L
distributions have a common parent distribution at a probability of
1.4$\times 10^{-10}$. M/L scales with relaxation time almost in the
same way as with mass: $\frac{{\rm d} log(M/L))}{{\rm d} log(t_{\rm
    relax})}=0.150 \pm 0.021$.

Apart from studying the relative difference in M/L ratio between GCs
and UCDs, it is also important to compare the M/L ratios with the
model predictions on an absolute scale. While the average M/L of UCDs
is $\sim$40\% {\it above} the 13 Gyr isochrone, M/L ratios of the
Galactic globular clusters are {\it below} the 13 Gyr isochrone by the
same factor.  The mean age derived from their location with respect to
the isochrones is 7-8 Gyrs, well below the typical globular cluster
age of $\sim$12-13 Gyrs, indicating that the input for the stellar
population codes may not represent the globular cluster properties (see also
Dabringhausen, Hilker \& Kroupa 2008). In what follows,
we therefore discuss the extent to which dynamical evolution
may have changed M/L ratios for the compact stellar systems
under investigation, and hence contributed to shaping the observed
trend between mass and M/L.

\subsubsection{Does dynamical evolution shape the M/L trend?}
Baumgardt \& Makino (2003) showed that star
clusters experience a depletion in low-mass stars leading to a drop in
M/L of up to 0.5 after about 0.8 dissolution timescales. This drop in
M/L corresponds to about 30\% of the mean measured M/L ratio of
Galactic globular clusters. Dynamical evolution could hence be
responsible for the lower M/L ratios of GCs, provided that their
dissolution timescale is comparable to or smaller than a Hubble time.
To estimate the dissolution timescale $t_{\rm \rm diss}$ for a typical
Galactic globular cluster, we apply equation (6) from the recent study
of Lamers et al. (2006), in which the photometric evolution of
dissolving star clusters in the Galaxy's gravitational potential is
investigated.
\begin{equation}
{\rm t_{\rm diss}}=6.60 \times 10^2(\frac{M_i}{10^4})^{0.653}\times t_0^{0.967-0.00825\times log(M_i/10^4)}
\label{tdiss}
\end{equation}

$M_i$ is the initial cluster mass. The time-scale $t_0$ depends on the
tidal field of the environment.  Lamers et al. (2006) adopt $t_0=$21.8
Myr, which is valid for a circular orbit in the Galaxy at 8.5 kpc
radial distance (Baumgardt \& Makino 2003).  The median galactocentric
distance of the Galactic GCs plotted in Fig.~\ref{mass_ML_all} is 9.2
kpc (McLaughlin \& van der Marel 2005), very close to the assumed 8.5
kpc. Therefore, we also adopt $t_0=$21.8 Myr for estimating $t_{\rm \rm
  diss}$. Evaluating equation ~\ref{tdiss}, we obtain $t_{\rm \rm diss}
\simeq$50 Gyrs for 10$^5$ M$_{\rm \sun}$, and $t_{\rm \rm diss} \simeq$250
Gyrs for 10$^6$ M$_{\rm \sun}$. That is, dynamical evolution should not
have changed the primordial M/L ratios of compact stellar systems with
masses above 10$^6$ M$_{\rm \sun}$, including the UCDs.  For Galactic GCs
in Fig.~\ref{tdiss}, masses are between 10$^4$ and 10$^6$ M$_{\rm \sun}$,
such that their dissolution timescales are closer to, and in some
cases below, a Hubble time. Assuming that the GC M/L ratio decreases
linearly up to a difference of 0.5 after 0.8 dissolution timescales
(Baumgardt \& Makino 2003), we plot in the bottom left panel of
Fig.~\ref{mass_ML_all} the expected M/L ratios of Galactic GCs if
there was no dynamical evolution. The corrections are small --- on
average about 5\%, which is neglibile in the context of this discussion.
Provided that the absolute scale of dissolution times derived by
Baumgardt \& Makino (2003) is applicable to the Milky Way GCs included
in this study, their M/L ratios should not have notably decreased due
to dynamical effects from their initial value. 

\vspace{0.15cm}

\noindent The stellar population
models used here indeed appear to over-estimate the M/L ratios of globular
clusters with a canonical IMF by $\sim$40\%, and on average under-estimate
the M/L ratios of UCDs by about the same amount.

\begin{figure}[]
\begin{center}
\epsfig{figure=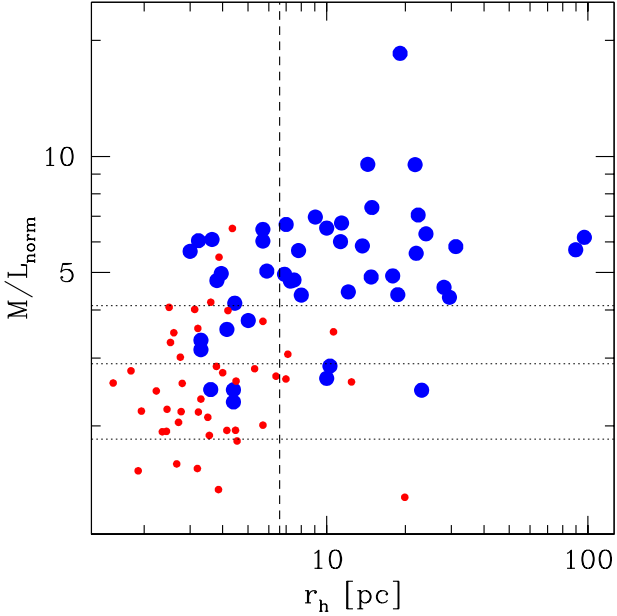,width=8.6cm}
\caption{Normalised M/L ratio of the sample of compact stellar systems from Fig.~\ref{mass_ML_all} plotted against their effective half-light radius $r_h$ in pc. Small (red) dots indicate GCs, defined as compact stellar systems with M$<2\times10^6$M$_{\rm \sun}$. Large (blue) dots are UCDs, defined as compact stellar systems with M$>2\times10^6$M$_{\rm \sun}$. The vertical dashed line indicates the radius
where the relaxation time is equal to one Hubble time.}
\label{rad_ML}
\end{center}
\end{figure}

\begin{figure}[]
\begin{center}
\epsfig{figure=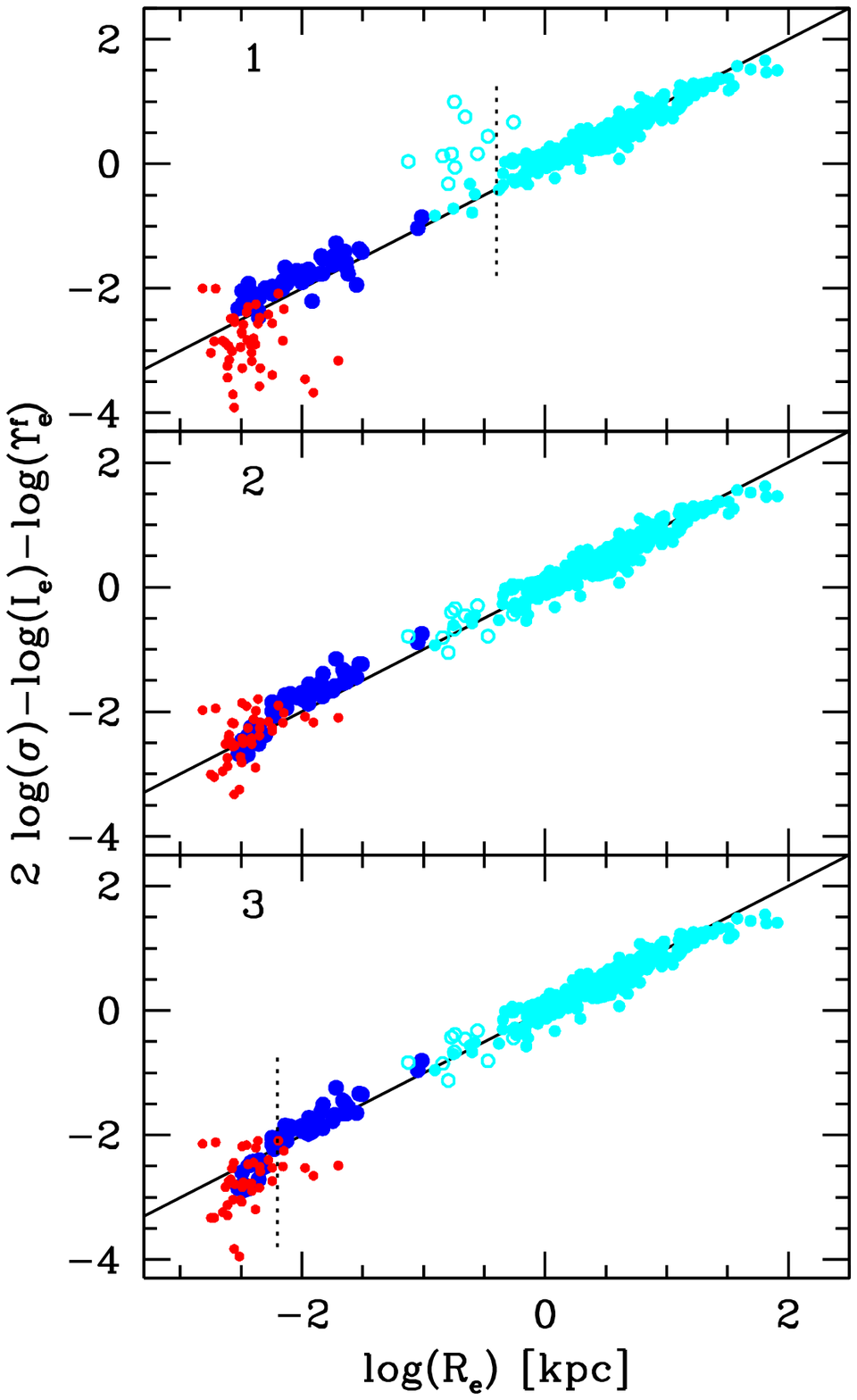,width=8.8cm}
\caption{The location of the compact stellar systems from Figs.~\ref{mass_ML_all} and~\ref{rad_ML} with respect to three different formulations of the 'fundamental manifold' (Zaritsky et al. 2006a, Zaritsky et al. 2006b). Table~\ref{manifold_coef} gives the fitting coefficients for the three formulations. Large blue dots are UCDs, small red dots are GCs. A global mass cut at 2$\times10^6$M$_{\rm \sun}$ is applied to separate UCDs from GCs. Cyan dots are large spheroids from the sample of Zaritsky et al. (2006a), with open cyan circles indicating faint Local Group dSphs with $M_V>-9$ mag (Zaritsky et al. 2006b, Simon \& Geha 2007). {\bf Plot 1:} This is the original formulation of the manifold (Zaritsky et al. 2006a), for which the fit of $log({\Upsilon}^{f}_{\rm e}(\rm log(\sigma),\rm log(\rm I_e))$ only includes large stellar spheroids with $log(r_e)>-0.4$, excluding the Local Group dSphs. {\bf Plot 2:} The fit of $log({\Upsilon}^{f}_{\rm e}(\rm log(\sigma),\rm log(\rm I_e))$ includes all objects in the plot (Zaritsky private communication).
  {\bf Plot 3:} The fit of $log({\Upsilon}^{f}_{\rm e}(\rm
  log(\sigma),\rm log(\rm I_e))$ includes all objects in the plot except
  UCDs with $\log(r_e)<-2.2$ and GCs (Zaritsky, private
  communication). See also Fig.~\ref{rad_ML} and text. The limiting
  $\log(r_e)$ is marked by a dashed vertical tick.}
\label{manifold}
\end{center}
\end{figure}
\subsection{Environmental dependence of M/L ratios}

From Fig.~\ref{mass_ML_all} it is evident that M/L ratio measurements
of the Fornax UCDs fit well into the general trend of M/L increasing with
mass. However, it is also interesting to note that we have not found
a Fornax UCD with such extraordinarily high M/L ratios as the three DGTOs
from Ha\c{s}egan et al. (2005). Fornax UCDs cover the same mass range as Virgo
UCDs, but their average M/L ratio is only 0.61 $\pm$ 0.11 that of the
Virgo UCDs, or 0.71 $\pm$ 0.08 when excluding the Virgo UCD with the
highest M/L ratio (S999). 

Such a M/L ratio difference may arise from age differences. If Fornax
UCDs have luminosity weighted ages around 7 Gyrs and Virgo UCDs ages
around a Hubble time, both populations would be equally {\it
  inconsistent} with M/L ratio predictions from stellar populations
for their age, indicating the presence of (baryonic or non-baryonic)
dark matter (Mieske \& Kroupa 2008, Dabringhausen, Hilker \& Kroupa
2008).  Derivation of spectroscopic ages from line abundances for
Virgo UCDs have shown that they are most consistent with old ages
around a Hubble time (Evstigneeva et al. 2007). For Fornax, the
situation is less clear.  Mieske et al. (2006) find indications for
intermediate ages in Fornax UCDs from relating H$_\beta$ to
metallicity sensitive line indices, but these data were not calibrated
to the Lick system.  Accurate age determinations for a comprehensive
sample of Fornax UCDs by spectroscopy or multi-band photometry will
allow to draw firmer conclusions in this context.

The lower M/L ratios of Fornax UCDs --- provided they are {\it not}
explained by age differences --- may support the assumption of
different competing formation channels for UCDs (Mieske et al.  2006). In that
context, the higher M/L ratios of Virgo UCDs could be interpreted as
being due to dark matter.  Goerdt et al. (2008) show that under
certain conditions, remnants of tidally stripped dwarf galaxies can
maintain a significant amount of dark matter (although see Bekki et
al.  2003 for a different view).  The lower M/L ratios in Fornax UCDs
could then be explained by them being stellar super clusters formed
without dark matter (Fellhauer \& Kroupa 2002, Mieske et al. 2006).

\subsection{Fundamental scaling relations of stellar systems: how do UCDs and GCs fit in?}

Here we analyse how compact stellar systems such as UCDs and GCs fit
into fundamental scaling relations for more extended stellar systems.
In motivating this analysis, we show in Fig.~\ref{rad_ML} the half-light radius of
the compact stellar systems from Fig.~\ref{mass_ML_all} plotted against
their M/L ratio. We mark the radius at which the relaxation time is
equal to one Hubble time, which is roughly 7 pc (log($r_h$)=0.82; or
log($r_h$/kpc)=$-$2.18). As also seen in Fig.~\ref{mass_ML_all}, this
limit nicely marks the rise of M/L ratios between the regime of GCs
and UCDs.

In Fig.~\ref{manifold} we show the location of all compact stellar
systems from Fig.~\ref{mass_ML_all} in the so-called 'fundamental
manifold' (Zaritsky et al. 2006a \& 2006b, Zaritsky et al. 2008). 
  The fundamental manifold concept aims at a unifying empirical
  description of the structural and kinematic properties of stellar
  spheroids. It relates the effective radius $r_e$ to velocity
dispersion $\sigma$ and effective I-band surface brightness $\rm I_e$.
In their studies, Zaritsky et al. show that stellar spheroids
  from the scale of galaxy clusters ($\sim 10^5$ pc) down to the scale
  of dwarf elliptical galaxies ($\sim 10^2$ pc) appear to form a
  common sequence in this manifold. With the data from
  Fig.~\ref{mass_ML_all} we can extend these considerations down to
  the smallest stellar systems ($\sim 10^0$ pc).

To derive $\rm I_e$ for the compact
stellar systems, we use (V-I) measurements where available, and
otherwise convert [Fe/H] to (V-I) using the calibration relation of
Kissler-Patig et al. (1998). In the plots we adopt a simple mass limit
of 2$\times 10^6$ M$_{\rm \odot}$ to separate UCDs from GCs (see
Fig.~\ref{mass_ML_all}).
  
Zaritsky et al. (2006a, 2006b, 2008) define the fundamental manifold relation
as 

\footnotesize
\begin{equation}
\label{mf1}
\rm log(r_e) = 2 \times log(\sigma) - log(\rm I_e) - log({\Upsilon}^{f}_{\rm e})
\end{equation}
\normalsize

In this formulation, $\rm log({\Upsilon}^{f}_{\rm e})$ is the effective
mass-to-light ratio parametrized in terms of $\rm log(\sigma)$ and
$\rm log(\rm I_e)$. That is, $\rm
log({\Upsilon}^{f}_{\rm e})=log({\Upsilon}^{f}_{\rm e}(\rm log(\sigma),\rm
log(\rm I_e))$. The parametrization is determined from a fit of
dynamically derived M/L ratio (using the virial theorem) as a function
of $\rm log(\sigma)$ and $\rm log(\rm I_e)$ (see Zaritsky et al. 2008). It
is clear that the exact functional shape of
$\rm log({\Upsilon}^{f}_{\rm e}(\rm log(\sigma),\rm log(\rm I_e))$ --- and hence
the location of the fundamental manifold --- depends on which stellar
systems are included in the fit.  For example, the original
formulation of the manifold (Zaritsky et al.  2006a) does not include
the heavily dark matter dominated Local Group dwarf spheroidal
galaxies in the fit (Zaritsky et al. 2006b, Simon \& Geha 2007). A
revised formulation extending to the very large M/L of the dSphs was
presented in Zaritsky et al. (2008; see Table 1 of that paper).

We are interested in the link between UCDs and canonical galaxies on
the one hand, and on the relation between UCDs and star clusters on
the other hand. Therefore, we show in Fig.~\ref{manifold} the location
of UCDs and GCs with respect to three different formulations of the
fundamental manifold. The functional form of $\rm
log({\Upsilon}^{f}_{\rm e}(\rm log(\sigma),\rm log(\rm I_e))$ for
these three representations is indicated in Table~\ref{manifold_coef}.

The first formulation does not include Local Group dwarf galaxies to
the fit of $\rm log({\Upsilon}^{f}_{\rm e}(\rm log(\sigma),\rm
log(\rm I_e))$, nor UCDs and GCs. This is the original manifold version
from Zaritsky et al.  (2006a).  It is intriguing that in this
formulation, {\it UCDs extend the fundamental manifold relation} by
more than a decade in $r_e$, down to $r_e \sim 5-7$ pc (note that $r_e$
is shown in units of kpc in Fig.~\ref{manifold}).  Together with all
other spheroids they follow a well defined linear function slightly
inclined with respect to the original fundamental manifold, with a
slope 0.92 $\pm$ 0.01.  {\it The fundamental manifold relation in this
  formulation breaks down only for the faintest dwarf spheroidal
  galaxies ($M_V>-9$ mag) and for globular clusters.} A possible
interpretation of this is that for the faintest dwarf spheroidals, the
dark matter halo is de-coupled from the baryons (see also Zaritsky et
al. 2006b) such that the continous relation of baryon packing
efficiency vs. galaxy scale breaks down.  Another possibility is that
the faintest dwarfs are out of dynamical equilibrium. This aspect is
closely related to the discussion of the origin of dwarf satellite
galaxies (dark-matter dominated cosmological substructure vs. tidal
dwarf galaxy, see Kroupa et al. 2005).

The second formulation includes all objects in Fig.~\ref{manifold} for
the fitting $\rm log({\Upsilon}^{f}_{\rm e}(\rm log(\sigma),\rm
log(\rm I_e))$ (Zaritsky, private communication). Again, UCDs follow the
manifold line, and only for $\rm log(r_e)\lesssim-2.2$ they start to
'bend down'.  Interestingly, at this radius also the transition
between objects with relaxation times smaller and larger than a Hubble
time occurs (Fig.~\ref{rad_ML}). Globular clusters show a large
scatter, and do clearly not align along the manifold.

The third formulation includes all objects for fitting {\it except}
GCs and UCDs with $\rm log(r_e)<-2.2$. This formulation hence excludes
dynamically relaxed stellar systems from the fit. UCDs with
$\rm log(r_e)>-2.2$ align very well with the manifold, while globular
clusters and smaller UCDs do not.

\vspace{0.2cm}

Summarizing, UCDs with $\rm log(r_e)>-2.2$ ($r_e\gtrsim$7 pc) appear
to form a single family with larger stellar systems in the fundamental
manifold. The location of most GCs is inconsistent with the
fundamental manifold extrapolated from larger stellar systems.
Dynamically un-relaxed stellar systems appear to form a single
manifold, while the relaxed systems --- due to their advanced dynamical
evolution --- scatter very broadly around it.

\begin{table*}
\caption{Fitting coefficients for $\rm log({\Upsilon}^{f}_{\rm e}(\rm log(\sigma),\rm
log(\rm I_e))$ in the three fundamental manifold formulations from Fig.~\ref{manifold}. The functional form is $\rm log({\Upsilon}^{f}_{\rm e})= c1 + c2* log(\sigma) + c3*log(\rm I_e) + c4*(log(\rm I_e))^2 + c5*(log(\sigma))^2 + c6*log(\sigma)*log(\rm I_e)$.}
\begin{center}
\begin{tabular}{|l|rrrrrr|}\hline
Manifold formulation & c1 & c2 & c3 & c4 & c5 & c6 \\\hline
1 & 2.75 & $-$1.70 &$-$0.295 & ----- & 0.63 & -----\\
2 & 1.8974 & 0.1896 & $-$0.9699 & 0.1095 & 0.1193 & 0.02893\\
3 & 2.2397 & $-$0.3006 &  $-$0.8726 &  0.1159 &  0.2827 & $-$0.0337\\\hline 
\end{tabular}
\end{center}
\label{manifold_coef}
\end{table*}

\section{Summary and conclusions}
\label{conclusions}
In this paper we have analysed the internal dynamics of 23
ultra-compact dwarf galaxies in the Fornax cluster. The analysis is
based on high-resolution spectroscopy obtained with the FLAMES
spectrograph at the VLT. Our targets cover an approximate mass
range of $10^6<M<10^8$ M$_{\sun}$ and a luminosity range
$-10.4<M_V<-13.5$ mag, overlapping the bright end of the globular
cluster luminosity function. We also compare the dynamical properties
of UCDs and GCs, and put them into the context of fundamental scaling
relations defined for larger galaxies. We obtain the following
results:

\begin{enumerate}
  
\item In the M$_V$-$\sigma$ plane, we find that UCDs with $M_V<-12$
  mag are consistent with the extrapolation of the
  Faber-Jackson relation for luminous elliptical galaxies. For $M_V>-12$ mag,
  most objects are located closer to the extrapolation to brighter
  luminosities of the globular cluster M$_V$-$\sigma$ relation.

\item We derive dynamical M/L ratios for those 15 of the 23 UCDs for
  which HST archival imaging is available, taking into account
  aperture effects in the spectroscopy (Hilker et al.  2007). Three out of
  the 15 UCDs have dynamical M/L ratios too high to be explained by
  canonical stellar populations, but we do not find Fornax UCDs with
  M/L ratios as extreme as found for some Virgo UCDs (Ha\c{s}egan et al.
  2005). At a given metallicity, Fornax UCDs have on average 30 to
  40\% lower M/L ratios than Virgo UCDs.

\item We normalise the dynamical M/L ratios of the 15 Fornax UCDs to
  solar metallicity, using predictions from stellar population models
  (Bruzual \& Charlot 2003, Maraston 2005). We find no significant
  correlation between normalised M/L ratio and mass or relaxation time
  for our Fornax UCD sample. We do not find a dependence of normalised
  M/L ratio on projected clustercentric distance.

\item We add our new measurements for 15 Fornax UCDs to the available
  data on M/L ratios of compact stellar systems in the broader mass
  range $10^4<M<10^8$M$_{\sun}$. We include Galactic globular clusters
  and UCDs in Virgo, CenA and Fornax. We re-analyse dynamical mass
  estimates of UCDs in the Virgo cluster (Ha\c{s}egan et al. 2005) and the
  CenA group (Rejkuba et al.  2007), using our modelling algorithm
  (Hilker et al.  2007) to correct for aperture effects in the
  spectroscopy. The corrections for the global velocity dispersion
  $\sigma$ are of the order of 5-10\%. We also provide previously
  unpublished M and M/L estimates for 5 CenA compact objects.
  
\item We find a clear break in the distribution of normalised M/L
  ratios at a characteristic mass of $\simeq$2$\times 10^6$M$_{\sun}$,
  which roughly corresponds to a relaxation time of one Hubble time
  (see also Dabringhausen, Hilker \& Kroupa 2008, Mieske \& Kroupa
  2008).  Objects more massive than this limit have normalised M/L
  ratios twice as large as objects less massive than this limit. In
  this context we suggest to separate UCDs from GCs by a mass limit of
  M$\simeq 2\times 10^6$M$_{\sun}$ (see also Ha\c{s}egan et al. 2005). On
  average, the M/L ratios of UCDs are 40\% above the expectations for
  a 13 Gyr stellar population with canonical IMF, while for GCs they
  are 40\% below these expectations.  We find that the M/L ratios
  estimates of GCs are probably only weakly biased ($\sim$5\%) by
  their dynamical evolution, indicating that stellar population models
  indeed over-predict M/L ratios for compact stellar systems like GCs.
  
\item UCDs extend the 'Fundamental Manifold' in its original
  formulation (Zaritsky et al.  2006a) by more than a decade in $r_e$
  down to $r_e \sim 5-7$ pc. In this formulation, neither the faintest
  dwarf spheroidals ($M_V>-9$ mag) nor GCs lie on the manifold. When
  using also GCs, UCDs and dSphs to define the shape of the FM, UCDs
  with $r_e \gtrsim 7$pc and dwarf spheroidals align along the
  manifold, while GCs and smaller UCDs do not. This characteristic
  scale of $r_e$$\simeq$7pc also marks the transition between compact
  stellar systems with relaxation times below and above a Hubble time.

\end{enumerate}

\noindent We suggest a defintion of UCDs as those compact stellar
systems with M$\ge$2$\times 10^6$M$_{\sun}$ and 7$\lesssim$${\rm
  r_e/pc}$$\lesssim$100.  As such, UCDs are the smallest dynamically
un-relaxed stellar systems. From their position in the 'Fundamental
Manifold' they can be considered the small-scale end of the galaxy
sequence.

A key question about UCDs is whether they are of
'cosmological' origin, hence related to compact low-mass dark matter
halos.  Their elevated M/L ratios can be interpreted as marking the
on-set of dark matter domination in small stellar systems.  However,
dark matter can hardly be detected directly, such that observational
efforts need to be directed towards verifying/excluding alternative
scenarios, such as a variation of the IMF in UCDs (Mieske \& Kroupa
2008). In parallel, theoretical
studies regarding the dynamical evolution of compact stellar systems
embedded in dark matter halos are needed for the mass-size regime of UCDs.

In this paper it has been found that Fornax UCDs have
30-40\% lower dynamical M/L ratios than Virgo UCDs.  A possible
explanation for this is that only Virgo UCDs have significant
fractions of dark matter. This may be explained by  the dominance of
different UCD formation channels in Virgo and Fornax (Mieske et al. 2006).
A simple way to test the possibility of different dark matter
fractions is to determine the luminosity weighted ages of Fornax and
Virgo UCDs.  Younger ages in Fornax UCDs of $\sim$7 Gyrs would
naturally explain the M/L ratio differences and imply similar dark
matter fractions as in Virgo.  Together with efforts to constrain the
IMF shape in UCDs, such an observational study is the next logical
step in understanding the puzzling nature of UCDs.

\acknowledgements

We warmly thank Dennis Zaritsky for providing us with updated fundamental
manifold formulations. We also thank Mat\'{i}as Gomez
for providing us with structural parameters for GCs in CenA.

\begin{longtable}{|l|rr|rr|rr|}
\caption{\label{table2} This table gives masses, M/L$_V$ ratios, metallicities, M/L$_V$ ratios normalised to solar metallicity, half-light radii and velocity dispersions $\sigma$ for the compact stellar systems from Fig.~\ref{mass_ML_all} to Fig.~\ref{manifold}. The table is ordered by descending mass. Sources are this paper$^a$, Hilker et al. (2007)$^b$, Rejkuba et al. (2007)$^c$, Ha\c{s}egan et al. (2005)$^d$, Evstigneeva et al. (2007)$^e$, McLaughlin \& van der Marel (2005)$^f$, Meylan et al. (2001)$^g$, Barmby et al. (2007)$^h$, de Marchi et al. (1999)$^i$. For galactic sources (ID MW....; McLaughlin \& van der Marel 2005), the quoted $\sigma$ measurements are close to central values.  For extragalactic sources, $\sigma$ is the global velocity dispersion, corrected for spectroscopy aperture losses (see text and Hilker et al. 2007). }\\
\hline ID & Mass [10$^6$M$_{\rm \sun}$] & M/L & [Fe/H] [dex] &
M/L$_{\rm \rm norm}$ & r$_{\rm \rm h}$ [pc] & $\sigma$ [km/s]\\\hline
\endfirsthead
\caption{continued.}\\
\hline
ID & Mass [10$^6$M$_{\rm \sun}$] & M/L & [Fe/H] [dex] & M/L$_{\rm \rm norm}$ & r$_{\rm \rm h}$ [pc] & $\sigma$ [km/s]\\
\hline
\endhead
\hline
\endfoot
F-19$^a$               &             93.6 (14.0)     &   4.69 (0.70)  &  -0.4 &  5.73 (1.03)  &   89.7 &  22.8\\
VUCD7$^e$              &             88.3 (22.0)     &   4.39 (1.10)  &  -0.7 &  6.17 (1.61)  &   96.8 &  27.2\\
VUCD3$^e$              &              40.0 (5.9)    &   4.35 (0.64) & 0.0 &  4.38 (0.83)  &   18.7 &  35.8\\
UCD1$^b$               &             32.1 (3.9)    &   4.99 (0.60)  &  -0.7 &  7.05 (1.02)  &   22.4 &  27.1\\
S417$^d$               &             29.5 (6.0)      &   6.68 (1.40)  &   -0.7 &  9.54 (2.08)  &   14.4 &  29.8\\
VUCD5$^e$              &             29.1 (4.3)    &   4.02 (0.60)  &  -0.4 &   4.90 (0.88) &   17.9 &  26.4\\
VUCD1$^e$              &             28.2 (4.7)    &   4.11 (0.69) &  -0.8 &  6.01 (1.11)  &   11.3 &  32.2\\
F-24$^a$               &             24.5 (7.8)    &   3.44 (1.10)  &  -0.4 &  4.31 (1.43)  &   29.5 &  21.4\\
VUCD4$^e$              &             24.3 (6.3)    &   3.45 (0.89) &    -1.0 &  5.61 (1.49)  &    22.0 &  21.3\\
S999$^d$               &             23.4 (4.3)    &   10.2 (1.90)  &   -1.4 &  18.5 (3.5)   &   19.1 &  22.7\\
S928$^d$               &             19.3 (4.5)    &   5.32 (1.20)  &   -1.3 &  9.52 (2.25)  &   21.8 &  19.1\\
UCD5$^b$               &             18.0 (4.5)    &   3.37 (0.85) &   -1.2 &  5.84 (1.50)   &   31.2 &  18.7\\
VUCD6$^e$              &             17.7 (5.5)    &   3.02 (0.94) &    -1.0 &  4.87 (1.54)  &   14.8 &  22.3\\
F-1$^a$                &             16.2 (3.8)    &   2.45 (0.58) & 0.0 &  2.48 (0.66) &   23.1 &  18.7\\
S490$^d$               &             14.5 (0.3)   &   6.81 (0.15) &   0.2 &  6.09 (0.82) &   3.6 &  41.6\\
F-9$^a$                &             14.1 (3.6)    &   4.72 (1.20)  &  -0.8 &  6.96 (1.85)  &   9.1 &  25.7\\
F-5$^a$                &             13.7 (2.4)    &   3.16 (0.55) &  -0.3 &  3.75 (0.76) &   5.0 &  34.5\\
F-6$^a$                &             12.5 (2.4)    &   5.32 (1.00)    &   0.2 &  4.75 (1.09)  &   7.3 &  27.3\\
HCH99-18$^c$           &             11.2 (4.3)    &   3.68 (1.40)  &  -1.0 &  5.86 (2.26)  &   13.7 &  18.7\\
F-7$^a$                &             10.5 (1.4)    &   4.21 (0.57) &   -1.3 &  7.37 (1.07)  &   14.9 &  20.1\\
S314$^d$               &             9.1 (1.3)    &   4.63 (0.68) &   -0.5 &  6.05 (1.04)  &   3.2 &  34.9\\
F-12$^a$               &             8.3 (2.9)    &   2.36 (0.83) &  -0.4 &  2.86 (1.05)  &   10.3 &  22.9\\
G1$^g$                 &             7.2 (1.2)    &    3.6 (0.60)  &  -1.0 &  5.67 (1.01)  &     3.0$^h$ &   25.0\\
HGHH92-C1$^c$          &             6.8 (1.7)    &   3.67 (0.90)  &   -1.2 &   6.30 (1.58)  &    24.0 &  11.1\\
HGHH92-C23$^c$         &             6.6 (2.2)    &   1.68 (0.55) &   -1.5 &  3.16 (1.04)  &    3.3 &  29.5\\
HGHH92-C7$^c$          &             6.3 (2.2)    &   2.68 (0.95) &   -1.3 &  4.78 (1.71)  &    7.5 &   19.1\\
F-17$^a$               &             6.3 (1.6)    &   2.22 (0.55) &  -0.8 &  3.34 (0.86) &    3.3 &  28.5\\
F-11$^a$               &             5.7 (3.7)    &   1.64 (1.10)  &  -0.9 &  2.49 (1.63)  &    3.6 &  26.2\\
HCH99-15$^c$           &             5.6 (1.7)    &   3.11 (0.95) &    -1.0 &  5.04 (1.57)  &    5.9 &  20.5\\
F-34$^a$               &             5.5 (1.3)    &   3.17 (0.74) &  -0.9 &  4.97 (1.20)   &  4.0 &  24.6\\
F-22$^a$               &             5.3 (1.0)   &   2.13 (0.39) &   -0.4 &  2.66 (0.55) &    10.0 &  22.8\\
HGHH92-C11$^c$         &             5.3 (1.9)    &   4.45 (1.60)  &  -0.5 &   5.70 (2.18)  &    7.8 &  17.1\\
HGHH92-C17$^c$         &             5.1 (1.7)    &   3.39 (1.10)  &   -1.3 &  6.03 (1.98)  &    5.7 &  19.8\\
VHH81-C5$^c$           &             5.0 (1.2)    &   3.39 (0.80)  &   -1.6 &  6.52 (1.57)  &    10.0 &  14.8\\
HGHH92-C21$^c$         &             4.8 (1.7)    &   3.87 (1.40)  &   -1.2 &  6.66 (2.35)  &     7.0 &  17.2\\
H8005$^d$              &             4.8 (2.5)    &   2.61 (1.40)  &   -1.3 &  4.58 (2.38)  &   28.1 &   8.5\\
HCH99-2$^c$            &             4.2 (1.6)    &   3.62 (1.40)  &   -1.5 &  6.72 (2.62)  &   11.4 &   12.5\\
F-53$^a$               &             3.9 (1.0)      &   2.66 (0.69) &  -0.9 &  4.16 (1.11)  &   4.4 &  19.6\\
HGHH92-C6$^c$          &             3.6 (0.9)    &    1.60 (0.40)  &  -0.9 &  2.48 (0.64) &    4.4 &   19.0\\
F-51$^a$               &             3.5 (0.9)   &   2.38 (0.62) &  -0.8 &  3.56 (0.96) &   4.2 &  20.1\\
HGHH92-C29$^c$         &             3.3 (1.1)    &   3.51 (1.20)  &  -0.7 &  4.95 (1.74)  &    6.9 &  14.5\\
$\omega$Cen$^c$        &             3.0 (0.5)    &    2.40 (0.40)  &   -1.6 &  4.61 (0.80) &     8.0$^i$ &   16.0\\
HGHH92-C22$^c$         &             2.6 (0.8)   &   2.76 (0.85) &   -1.2 &  4.77 (1.49)  &    3.8 &  17.2\\
VHH81-C3$^c$           &             2.4 (0.7)   &   1.68 (0.50)  &  -0.6 &  2.31 (0.72) &    4.4 &  15.2\\
HCH99-16$^c$           &             2.0 (0.7)   &   2.16 (0.80)  &   -1.9 &  4.45 (1.67)  &   12.1 &   8.4\\
HGHH92-C44$^c$         &             1.9 (0.6)   &   3.38 (1.10)  &   -1.6 &  6.47 (2.13)  &    5.7 &  12.1\\
HGHH92-C36=R01-113$^c$ &             1.8 (0.6)   &   2.25 (0.75) &   -1.5 &  4.19 (1.41)  &    3.6 &  14.7\\
HCH99-21$^c$           &             1.6 (0.9)   &   1.48 (0.85) &    -2.0 &  3.07 (1.77)  &    7.1 &  9.7\\
HHH86-C18$^c$          &             1.6 (0.5)    &   0.95 (0.30)  &   -1.1 &  1.55 (0.50) &    3.2 &  14.7\\
MW-NGC6715$^f$         &             1.5 (0.6)    &   1.41 (0.58) &   -1.6 &  2.69 (1.11)  &    6.4 &  14.2\\
F-59$^a$               &             1.3 (0.6)   &  0.94 (0.43) &   -2.1 &  2.01 (0.93) &    5.7 &   9.8\\
HHH86-C15=R01-226$^c$  &             1.3 (0.5)   &   1.93 (0.75) &  -0.8 &  2.81 (1.11)  &    5.3 &   10.1\\
HGHH92-C41$^c$         &             1.2 (0.4)   &   1.87 (0.60)  &  -0.7 &  2.61 (0.87) &    4.5 &  10.7\\
R01-223$^c$            &             1.1 (0.4)    &   2.08 (0.75) &   -1.1 &  3.49 (1.27)  &    2.6 &  13.7\\
HHH86-C38=R01-123$^c$  &              1.1 (0.4)   &   1.48 (0.50)  &   -1.2 &  2.58 (0.88) &    2.8 &  13.6\\
HGHH92-C37=R01-116$^c$ &              1.1 (0.4)   &   1.49 (0.50)  &  -1.0 &  2.35 (0.80) &    3.3 &  12.0\\
MW-NGC6441$^f$         &            0.91 (0.47)   &   1.65 (0.85) &  -0.5 &  2.19 (1.14)  &   2.0 &   18.0\\
R01-261$^c$            &            0.87 (0.32)   &   0.96 (0.35) &  -1.0 &  1.53 (0.57) &    1.9 &  14.2\\
MW-NGC6388$^f$         &            0.79 (0.37)   &   1.89 (0.90)  &   -0.6 &  2.58 (1.24)  &   1.5 &  18.9\\
MW-NGC5824$^f$         &            0.68 (0.28)   &   1.96 (0.82) &   -1.9 &  3.98 (1.68)  &   4.2 &  11.6\\
MW-NGC104$^f$          &            0.64 (0.26)   &   1.33 (0.54) &  -0.8 &  1.95 (0.80) &   4.2 &  11.5\\
MW-NGC2808$^f$         &            0.61 (0.27)   &   1.46 (0.64) &   -1.2 &  2.47 (1.09)  &   2.2 &  13.4\\
MW-NGC6656$^f$         &            0.40 (0.19)   &   2.07 (0.99) &   -1.6 &  4.01 (1.92)  &   3.1 &    9.0\\
MW-NGC6864$^f$         &            0.40 (0.17)   &   1.78 (0.75) &   -1.2 &  3.02 (1.27)  &   2.8 &  10.3\\
MW-NGC6402$^f$         &            0.38 (0.21)   &   1.16 (0.64) &   -1.4 &  2.11 (1.16)  &   3.5 &   8.2\\
MW-NGC7089$^f$         &            0.36 (0.15)   &   0.98 (0.40)  &   -1.6 &  1.89 (0.78) &   3.6 &   8.2\\
MW-NGC6205$^f$         &            0.29 (0.12)   &   1.51 (0.62) &   -1.5 &  2.85 (1.19)  &   3.8 &   7.1\\
MW-NGC2419$^f$         &            0.29 (0.11)   &   0.61 (0.24) &   -2.1 &  1.31 (0.53) &   19.9 &    3.0\\
MW-NGC5272$^f$         &            0.28 (0.11)   &   1.39 (0.57) &   -1.6 &  2.65 (1.08)  &   7.0 &   5.6\\
MW-NGC1851$^f$         &            0.26 (0.10)    &   1.61 (0.65) &   -1.2 &  2.78 (1.12)  &   1.8 &  10.4\\
MW-NGC5286$^f$         &            0.24 (0.11)   &   0.99 (0.44) &   -1.7 &  1.93 (0.87) &   2.4 &    8.0\\
MW-NGC5904$^f$         &            0.18 (0.07)  &   0.78 (0.32) &   -1.3 &  1.37 (0.56) &   3.9 &   5.7\\
MW-NGC6254$^f$         &            0.17 (0.07)  &   2.16 (0.98) &   -1.5 &  4.06 (1.84)  &   2.5 &   6.6\\
MW-NGC3201$^f$         &            0.17 (0.07)   &   2.87 (1.20)  &   -1.6 &  5.48 (2.30)   &   3.9 &   5.2\\
MW-NGC6809$^f$         &            0.17 (0.07)  &   3.23 (1.30)  &   -1.8 &  6.51 (2.65)  &   4.4 &   4.9\\
MW-NGC5694$^f$         &            0.16 (0.06)  &   1.35 (0.54) &   -1.9 &  2.75 (1.11)  &     4.0 &   5.5\\
MW-NGC6341$^f$         &            0.12 (0.05)  &   0.88 (0.37) &   -2.3 &  1.94 (0.83) &   2.4 &   5.9\\
MW-NGC1904$^f$         &           0.096 (0.039)  &   1.16 (0.47) &   -1.6 &  2.21 (0.90) &   2.5 &   5.2\\
MW-NGC6171$^f$         &           0.084 (0.040)   &    2.20 (1.00)    &    -1.0 &  3.58 (1.71)  &   3.2 &   4.1\\
MW-NGC6218$^f$         &           0.084 (0.035)  &   1.77 (0.74) &   -1.5 &  3.29 (1.39)  &   2.5 &   4.5\\
MW-NGC6779$^f$         &           0.081 (0.035)  &   1.05 (0.45) &   -1.9 &  2.17 (0.94) &   3.2 &    4.0\\
MW-NGC6712$^f$         &           0.080 (0.039)  &   0.99 (0.48) &    -1.0 &  1.59 (0.77) &   2.7 &   4.3\\
MW-NGC288$^f$          &            0.078 (0.032)  &   2.15 (0.89) &   -1.2 &  3.74 (1.56)  &    5.7 &  2.91\\
MW-NGC6121$^f$         &           0.073 (0.035)  &   1.27 (0.61) &   -1.2 &  2.18 (1.06)  &   2.8 &   4.2\\
MW-NGC6362$^f$         &           0.058 (0.024)  &   1.16 (0.47) &  -1.0 &  1.83 (0.75)  &   4.5 &   2.8\\
MW-NGC5466$^f$         &           0.049 (0.020)   &   1.61 (0.67) &   -2.2 &  3.51 (1.49)  &   10.6 &   1.7\\
MW-NGC4590$^f$         &           0.044 (0.018)  &   0.92 (0.37) &   -2.1 &  1.95 (0.80) &   4.5 &   2.5\\
MW-NGC5053$^f$         &           0.038 (0.016)  &   1.18 (0.48) &   -2.3 &   2.60 (1.09)  &   12.4 &   1.4\\
MW-NGC4147$^f$         &           0.025 (0.010)   &   1.01 (0.42) &   -1.8 &  2.04 (0.86) &   2.7 &   2.6\\
MW-NGC6366$^f$         &          0.00807 (0.0044) &    0.30 (0.17) &  -0.8 &  0.45 (0.25)  &   3.1 &   1.3\\\hline
\end{longtable}
\end{document}